%% file: main.tex
\documentclass[lettersize,journal]{IEEEtran}
\usepackage{amsmath,amsfonts}
\usepackage{algorithmic}
\usepackage{algorithm}

\usepackage{array}
\ifCLASSOPTIONcompsoc
    \usepackage[caption=false, font=normalsize, labelfont=sf, textfont=sf]{subfig}
\else
\usepackage[caption=false, font=footnotesize]{subfig}
\fi
\usepackage{textcomp}
\usepackage{stfloats}
\usepackage{url}
\usepackage{verbatim}
\usepackage{graphicx}
\usepackage{cite}
\usepackage[]{hyperref}
\usepackage{tikz}
\usetikzlibrary{quantikz}
\usetikzlibrary{backgrounds}
\usepackage{float}

\def\BibTeX{{\rm B\kern-.05em{\sc i\kern-.025em b}\kern-.08em
    T\kern-.1667em\lower.7ex\hbox{E}\kern-.125emX}}

\begin{document}

\title{Domain-Specific Quantum Architecture Optimization}

\author{Wan-Hsuan Lin, Bochen Tan, Murphy Yuezhen Niu, Jason Kimko, and Jason Cong~\IEEEmembership{Fellow,~IEEE}

\thanks{W.-H.~Lin, B.~Tan, J.~Kimko, and J.~Cong are with the Department of Computer Science, University of California, Los Angeles, CA, USA; email: \{wanhsuanlin,bochentan,kimko\}@ucla.edu, cong@cs.ucla.edu.}
\thanks{M.~Y.~Niu is with Google AI Quantum, Venice, CA, USA; email: murphyniu@google.com.}
}


\maketitle

\begin{abstract}
With the steady progress in quantum computing over recent years, roadmaps for upscaling quantum processors have relied heavily on the targeted qubit architectures.
So far, similarly to the early age of classical computing, these designs have been crafted by human experts.
These general-purpose architectures, however, leave room for customization and optimization, especially when targeting popular near-term QC applications.
In classical computing, customized architectures have demonstrated significant performance and energy efficiency gains over general-purpose counterparts.
In this paper, we present a framework for optimizing quantum architectures, specifically through customizing qubit connectivity.
It is the first work that (1) provides performance guarantees by integrating architecture optimization with an optimal compiler, (2) evaluates the impact of connectivity customization under a realistic crosstalk error model, and (3) benchmarks on realistic circuits of near-term interest, such as the quantum approximate optimization algorithm (QAOA) and quantum convolutional neural network (QCNN).
We demonstrate up to 59\% fidelity improvement in simulation by optimizing the heavy-hexagon architecture for QAOA circuits, and up to 14\% improvement on the grid architecture.
For the QCNN circuit, architecture optimization improves fidelity by 11\% on the heavy-hexagon architecture and 605\% on the grid architecture.
\end{abstract}


\input{1_introduction}

\input{2_background}
\input{3_benchmark}

\input{4_architecture_space}
\input{5_search_strategy}
\input{6_evaluation_metrics}
\input{7_evaluation_results}
\input{8_conclusion}

\section*{Acknowledgment}
This work is partially supported by contributions from multiple industrial sponsors, including Intel, NEC, and Synopsys.

\bibliographystyle{IEEEtran}
\bibliography{bib/qas}

\newpage
\input{biography}

\end{document}

%% file: 1_introduction.tex
\section{ Introduction}
\label{sec:introduction} 

With state-of-the-art quantum computers reaching hundreds of qubits in size, we are at a historical turning point of moving beyond noisy intermediate--scale quantum (NISQ) devices towards scalable and error-corrected quantum computers.
Existing quantum architecture designs are based on theoretical quantum error correction research, which historically has been detached from the exciting progress in hardware engineering. 
As a result, the interface between quantum computing software and hardware, i.e., quantum computing architecture, is far from optimal and provides many opportunities for improvement.
Quantum computing architectures define which operations can be performed on certain qubits and how qubits are connected to one another.
The latter is represented by coupling graphs, in which the vertices are physical qubits and the edges are connections.
Two-qubit entangling operations can only occur between two adjacent qubits in the coupling graph.
To our knowledge, current quantum computing architectures have been designed exclusively by human experts and are not tailored to specific quantum applications.
Thus, many applications, e.g., those in quantum machine learning and optimization, are not optimally implemented on existing laboratory quantum computers in terms of required quantum circuit depth and circuit fidelity.
In pursuit of discovering the compelling commercial application of quantum computers and uncovering new quantum architectures that promise transformative advantage in scalability, it is necessary to improve and automate how we design quantum architectures.

Domain-specific architecture designs have led to some of the biggest accelerations in classical computing, including deep learning (e.g.,~\cite{iccad16-zhang-fang-zhou-pan-cong-caffine-dl-acceleration, tpu,iccad18-wei-liang-li-yu-zhang-cong-tile-architecture-cnn, isca18-deep-learning-acceleration, micro19-nvidia-simba-dl-architecture, fpga20-sohrabizadeh-wang-cong-optimization-deep-learning}),
large-scale genome analysis (e.g., \cite{asplos18-turakhia-bejerano-dally-darwin-genomics-fpga, fccm19-guo-lau-ruan-wei-cong-acceleration-genome-sequencing, isca20-genomic-acceleration, fccm20-lo-fang-wang-zhou-chang-cong-bqsr-accerlation-genome}), and
 big data applications (e.g.,  \cite{asplos2014-ouyang-sdf,isca2014-putnam-reconfigurable-accelerate-datacenter,islped2016-ouyang-extending-moore-new-data-center,micro2016-caulfield-cloud-scale-acceleration-architecture,isca20-samardzic-qiao-aggarwal-chang-cong-bonsai-sorting-fpga}).
By harnessing distinct features of each application domain, the computational capacity of the underlying hardware can be fully utilized to achieve what is otherwise impossible with a general-purpose architecture.
To obtain a similar gain for quantum computation, 
an effective design methodology must discover architectures that maximize circuit fidelity under realistic error modeling, thereby improving the success rate of the targeted algorithm.
When this success rate exceeds a certain threshold, a quantum computer has the potential to outperform a classical one \cite{nature19-google-quantum-supremacy}.

Existing efforts in domain-specific architecture search have yet to be successfully applied to quantum applications due to the large dimensionality of the search space as well as the difficulty in architecture evaluation due to a lack of realistic circuit performance estimation metrics.  
The study in \cite{date20-deb-dueck-wille-alternative-quantum-architectures} considers the potential of coupling graph modifications based on a given circuit's properties.
However, this work does not provide performance guarantees.
This heuristic approach leads to an inaccurate estimation of the architecture performance, and thus, may lead to a sub-optimal architecture.
The recent work in \cite{isca19-murali-linke-martonisi-abhari-nguyen-alderete-triq-architecture-studies} makes a valuable comparison between different architectures and presents some advantages of all-to-all connectivity for trapped ions.
However, the main focus is on comparison rather than designing new quantum architectures for domain-specific applications.
Li~\emph{et al.}~\cite{asplos20-li-ding-xie-superconducting-architecture-design} is the first work that proposes the concept of quantum architecture customization targeting on fixed-frequency superconducting qubits.
They present a hardware-design workﬂow for quantum applications to achieve high yield rates.
However, since their method adopts heuristic strategies for architecture optimization, 
and it is hard to guarantee the optimality of the resulting architecture.
In addition, they use post-mapping gate count as the evaluation metric, which may not accurately reflect the circuit performance due to the lack of consideration for other noise sources, e.g., qubit idling errors and crosstalk.

In this work, we propose the first domain-specific quantum architecture optimization framework to improve circuit fidelity for quantum applications to achieve guaranteed yield rates.
To overcome the inherently difficult search problem, we integrate an optimal compilation process with the architecture optimization procedure in a combined SMT formulation to enable optimal compilation results and offer a provable performance guarantee for the search outcome by modeling the architecture optimization as a constraint satisfaction problem.
Under this framework, we propose an algorithm that searches for the optimal architectures using different hardware requirements from a defined architecture space in order to minimize the required computational resources for a target circuit.
We demonstrate the effectiveness of customized architectures with a realistic circuit fidelity model and achieve 59\% fidelity improvement on heavy-hexagon architectures and 14\% fidelity improvement on grid architectures for QAOA circuits.
For QCNN circuits, we achieve 11\% fidelity improvement on heavy-hexagon architectures and 605\% fidelity improvement on grid architectures.
With our methodology, heavy-hexagon architectures, designed for scalable quantum error correction code, can achieve comparable circuit fidelity with grid architectures, despite having 33.3\% fewer degrees of connectivity.
Our work opens up new directions in quantum computing research towards building large-scale quantum computers by enabling a seamless integration of new quantum computing and error correction paradigms into the ever-changing quantum hardware engineering and application discovery landscapes through automated quantum architecture search.

The remainder of this paper is organized as follows.
Section~\ref{sec:background} covers background knowledge for quantum circuits and quantum circuit synthesis, and
Section~\ref{sec:nisq} introduces the NISQ applications used for evaluation.
Section~\ref{sec:architecture} defines our architecture space, 
Section~\ref{sec:qas} demonstrates our domain-specific architecture optimization methodology, and
Section~\ref{sec:eval_m} presents our crosstalk error model and the metrics for architecture performance evaluation.
Finally, we show the experimental results in Section~\ref{sec:eval} and discuss some future research directions in Section~\ref{sec:future}.

%% file: 2_background.tex
\section{Background}
\label{sec:background}

\subsection{Qubits, Quantum Gates, and Quantum Circuits}
The state of a qubit can be represented by the linear combination $a|0\rangle+b|1\rangle$, where $|0\rangle$ and $|1\rangle$ are basis quantum states in Dirac notation and $a$ and $b$ are complex coefficients.
Each additional qubit doubles the dimension of the quantum state space, so $n$ qubits can be in a superposition state of $2^n$ basis states from $|0\rangle$ to $|2^n\rangle$.
Quantum gates are unitary matrices that are applied on quantum states via matrix multiplication.
Measuring a qubit is an irreversible operation that collapses the qubit's superposition to extract state information.
The measurement outcome of a single qubit is $|0\rangle$ with probability $|\alpha|^2$ and $|1\rangle$ with probability $|\beta|^2$.
A quantum \emph{circuit} is a sequence of operations, including qubit initialization, gates acting on the qubits, and qubit measurements.
Due to environment noises and imperfect qubit operations, the success rate of a quantum circuit is measured by its circuit fidelity, which is affected by the circuit execution time, the total gate count, and the layout synthesis solution to be discussed next.

\subsection{Logic and Layout Synthesis}
To execute a quantum circuit on a quantum computer, 
logic and layout synthesis are needed to map the circuit onto the hardware architecture.
During logic synthesis, we translates the original gates in the circuit into supported gates from the target computer's gate set by gate decomposition.
For example, we can apply KAK decomposition~\cite{arxiv0507-tucci-kak-decomposition} to translate an arbitrary two-qubit gate into three CNOT gates and some single-qubit gates, which are supported by superconducting qubit architectures~\cite{nature19-google-quantum-supremacy,misc-ibm-quantum-experience}
At this point, any two-qubit gate can logically act on an arbitrary pair of qubits in the quantum circuit, but the physical execution of this gate requires the two involved qubits to be coupled in the target architecture.
Layout synthesis \cite{tc20-tan-cong-optimality-layout-queko}, also called qubit mapping~\cite{tcad08-maslov-falconer-mosca-placement, cgo18-siraichi-santos-collange-pereira-qubit-allocation} addresses this issue by mapping the logical qubits to physical qubits for each time step and scheduling gate operations while satisfying a given architecture's connectivity constraints as defined in its coupling graph.
Depending on the architecture, some two-qubit gates may still be scheduled on two non-adjacent qubits.
In this case, layout synthesizer can insert SWAP gates as needed to exchange the quantum states of two physical qubits, or directly schedule a long-range gate with some overhead, e.g., a bridge gate \cite{cgo18-siraichi-santos-collange-pereira-qubit-allocation}.
However, both approaches increase the total gate count and circuit depth, which increases the execution time and introduces the opportunity of more errors.

Layout synthesis is critical for evaluating architecture.
On the hardware side, we usually fix native gate sets since they are mainly determined by fundamental physical properties.
In comparison to the native gate sets, a wide range of coupling graphs has been introduced, e.g., \cite{misc-ibm-quantum-experience,prx20-chamberland-zhu-yoder-hertzberg-cross-codes-low-degree}.
As a results, an effective layout synthesizer needs to be adaptable to different qubit connectivities.
On the software side, logic synthesis already has established canonical and efficient methods. 
For instance, the common single-qubit and two-qubit gates appearing in many NISQ applications have a canonical decomposition to native gate sets \cite{arxiv0507-tucci-kak-decomposition}.
The layout synthesis problem, on the other hand, is NP-hard~\cite{tcad08-maslov-falconer-mosca-placement, cgo18-siraichi-santos-collange-pereira-qubit-allocation,tc20-tan-cong-optimality-layout-queko} with many heuristic approaches remaining far from optimal \cite{tc20-tan-cong-optimality-layout-queko}.
Moreover, the architecture selection heavily impact the results of layout synthesis and vice versa.

%% file: 3_benchmark.tex
\section{NISQ Applications}
\label{sec:nisq}

Circuit-based quantum computation realizes universal quantum operations by executing quantum circuits consisting of gates from a universal quantum gate set. 
The majority of NISQ applications utilizes the circuit-based model and chooses the universal gate set to be a set of single-qubit gates plus a two-qubit gate. 
This is also adopted in the research regarding quantum circuit complexity, e.g., Ref.~\cite{arxiv2106-haferkamp-faist-kothakonda-eisert-halpern-linear-complexity-growth,bremner2011classical,boixo2018characterizing}.
As mentioned in Section~\ref{sec:background}, a real-world quantum computation needs to adhere to connectivity constraints imposed by a quantum computer's coupling graph, resulting in a variety of circuit sizes depending on the optimality of the inserted gates.
In this study, in order to understand the potential of QC architecture customization, we focus on two NISQ applications: Quantum Approximate Optimization Algorithm~(QAOA)~\cite{arxiv1411-farhi-goldstone-gutmann-qaoa, natphys21-google-qaoa} and Quantum Convolutional Neural Network~(QCNN)~\cite{natphys19-cong-choi-lukin-quantum-convolutional-neural}.
For the concept of illustration, we choose QAOA due to its wide range of application for solving satisfiability optimization problems that has immediate practical use and QCNN for its valuable application on classifying many-body quantum states.
Additionally, as the goal of architecture optimization is to reduce the compilation overheard, these applications are chosen due to their demanding requirements on qubit connectivity.

\begin{figure}[!t]
    \centering
    \subfloat[QAOA algorithm involves executing the QAOA quantum circuit and classically optimizing the parameters in the circuit.]{
    \label{fig:qaoa-circuit}
        \begin{tikzpicture}
            \node[scale=0.65] (circuit) {
            \begin{quantikz}[row sep=3pt, transparent]
                &\lstick{$|0\rangle$} &\gate{H} &\gate[4]{U(C,\gamma_1)} &\gate{e^{-i\beta_1 X}} & \ \ldots\ \qw &\gate[4]{U(C,\gamma_p)} &\gate{e^{-i\beta_p X}} &\meter{}\rstick[wires=4]{}\\
                &\lstick{$|0\rangle$} &\gate{H} & &\gate{e^{-i\beta_1 X}} & \ \ldots\ \qw & &\gate{e^{-i\beta_p X}} &\meter{}\\
                &\lstick{$|0\rangle$} &\gate{H} & &\gate{e^{-i\beta_1 X}} & \ \ldots\ \qw & &\gate{e^{-i\beta_p X}} &\meter{}\\
                &\lstick{$|0\rangle$} &\gate{H} & &\gate{e^{-i\beta_1 X}} & \ \ldots\ \qw & &\gate{e^{-i\beta_p X}} &\meter{}\\
            \end{quantikz}
        };
        \node[scale=0.7, draw, yshift=-20pt] (optimizer) at (circuit.south) {Classical Optimizer};
        \draw[->, thick] (circuit.east) |- (optimizer.east) ;
        \draw[->, thick] (optimizer.north) -- ++(-20pt, 15pt)node[scale=0.9, pos=0.73, right]{$\beta_1$};
        \draw[->, thick] (optimizer.north) -- ++(30pt, 15pt)node[scale=0.9, pos=0.5, above]{$\gamma_p$};
        \draw[->, thick] (optimizer.north) -- ++(-60pt, 15pt)node[scale=0.9, pos=0.6, below]{$\gamma_1$};
        \draw[->, thick] (optimizer.north) -- ++(70pt, 15pt)node[scale=0.9, pos=0.6, below]{$\beta_p$};
        \end{tikzpicture}
    }
    \vfill
    \subfloat[Phase-splitting operator induced by a MAXCUT instance. (If a wire goes through a gate, then the gate does not operate on that wire, i.e., qubit.)]{
    \label{fig:phase-splitting}
        \begin{tikzpicture}
            \node[scale=0.65] (circuit) {
            \begin{quantikz}[row sep=3pt, transparent]
                &\gate[4]{U(C,\gamma)} &\qw & &\gate[2]{e^{-i\gamma Z Z}} &\qw &\qw &\gate[3, label style={yshift=-4pt}]{e^{-i\gamma Z Z}} &\gate[4]{e^{-i\gamma Z Z}}  &\qw\\
                & &\qw & & &\gate[2]{e^{-i\gamma Z Z}} &\gate[3, label style={yshift=8pt}]{e^{-i\gamma Z Z}} &\linethrough &\linethrough  &\qw\\
                & &\qw & &\gate[2]{e^{-i\gamma Z Z}} & &\linethrough & &\linethrough  &\qw\\
                & &\qw & & &\qw & &\qw & &\qw\\
            \end{quantikz}
        };
        \node[xshift=60pt] at (circuit.west) {=};
        \end{tikzpicture}
    }
    \caption{Applying QAOA to the MAXCUT problem.}
    \label{fig:qaoa}
\end{figure}
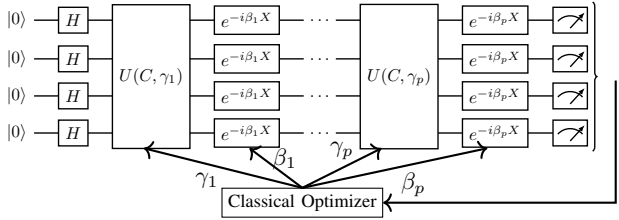
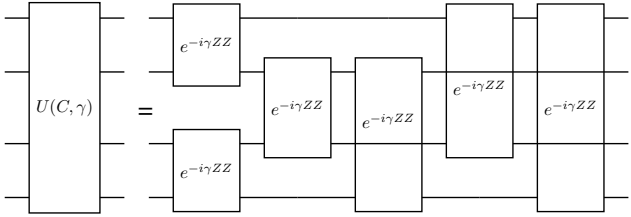

\noindent \textbf{Quantum Approximate Optimization Algorithm (QAOA).}
Many computational problems can be formulated as a quadratic unconstrained binary optimization (QUBO) problem~\cite{glover2018tutorial_qubo} or clauses of boolean variables in conjunctive normal form (CNF) which optimization objective is defined as the number of satisfied clauses.
QAOA \cite{arxiv1411-farhi-goldstone-gutmann-qaoa, natphys21-google-qaoa} aims to solve these optimization problems approximately with a quantum circuit consisting of $2p$ groups of gates 
\begin{equation}
\label{eq:qaoa_ansatz}
    U(B,\beta_p) U(C,\gamma_p) ... U(B,\beta_1) U(C,\gamma_1)|s\rangle,
\end{equation}
where $|s\rangle$ is the initialized superposition state produced by applying the Hadamard gate on all qubits at $|0\rangle$, and $\beta_1,\ldots,\beta_p$ and $\gamma_1,\ldots,\gamma_p$ are parameters.
Fig.~\ref{fig:qaoa-circuit} illustrates the QAOA circuit for the MAXCUT problem (NP-hard) on a graph.
The {\em mixing operator} $U(B,\beta_j)$ is a layer of single-qubit gates $e^{-i\beta_j X}$ on all qubits.
The {\em phase-splitting operator} $U(C,\gamma_j)$ consists of two-qubit gates $e^{-i\gamma_j Z_k Z_l}$ acting on all qubit pairs $(k,l)$, where each pair corresponds to an edge in the graph.
Fig.~\ref{fig:phase-splitting} shows the phase-splitting operator for a four-vertex complete graph where $e^{-i\gamma Z Z}$ gates are applied on all qubit pairs.
After alternatively applying mixing and phase-splitting operators $p$ times to the state $|s\rangle$, we measure the resulting quantum state.
According to the measurement results, classical optimizers derive $\beta$ and $\gamma$ parameters for the next iteration, as demonstrated in Fig.~\ref{fig:qaoa-circuit}.
The required connectivity for QAOA is decided by the graph derived from the input problem because the graph defines the phase-splitting operators~\cite{herrman2021qaoa_depth}.
Thus, QAOA often presents a challenging compilation problem when we map a non-planar graph to architectures laid across a 2D surface since non-local two-qubit gates are essential~\cite{natphys21-google-qaoa}. 

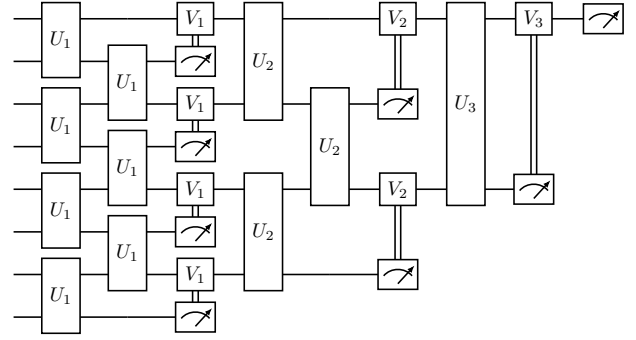
\begin{figure}
    \centering
    \begin{tikzpicture}
    \node[scale=0.75] {
    \begin{quantikz}[row sep=3pt]
        &\gate[2]{U_1} &\qw&\gate{V_1}        &\gate[3, nwires={2}]{U_2} &\qw                       &\gate{V_2}        &\gate[5, nwires={2, 3, 4}]{U_3} &\gate{V_3}&\meter{} \\
        &              &\gate[2]{U_1} &\meter{} \vcw{-1} &                          &                          &                  &                                &                 &\\
        &\gate[2]{U_1} &              &\gate{V_1}        &                          &\gate[3, nwires={2}]{U_2} &\meter{} \vcw{-2} &                                &                 &\\
        &              &\gate[2]{U_1} &\meter{} \vcw{-1} &                          &                          &                  &                                &                 &\\
        &\gate[2]{U_1} &              &\gate{V_1}        &\gate[3, nwires={2}]{U_2} &                          &\gate{V_2}        &                                &\meter{} \vcw{-4}&\\
        &              &\gate[2]{U_1} &\meter{} \vcw{-1} &                          &                          &                  &                                &                 &\\
        &\gate[2]{U_1} &              &\gate{V_1}        &                          &\qw                       &\meter{} \vcw{-2} &                                &                 &\\
        &              &\qw           &\meter{} \vcw{-1} &                          &                          &                  &                                &                 &\\
    \end{quantikz}
    };
    \end{tikzpicture}
    \caption{Quantum convolutional neural network on 8 qubits.
    The convolutional layers consist of the generic two-qubit unitary gates $U_i$ to be trained.
    The pooling layers consist of the single-qubit gates $V_j$ controlled by measurement results.}
    \label{fig:qcnn}
\end{figure}

\noindent \textbf{Quantum Convolutional Neural Network (QCNN).}
Classifying many-body quantum states is a valuable and computationally intensive problem in theoretical physics.
QCNN implements a quantum neural network that is NISQ-friendly and avoids the common vanishing gradient problem during training \cite{natphys19-cong-choi-lukin-quantum-convolutional-neural,prx21-arthur-qcnn-barren-plateau}.  
Fig.~\ref{fig:qcnn} displays an 8-qubit QCNN, where $U_i$ and $V_j$ are generic two- and single-qubit gates and the meter with two vertical lines represents a measurement outcome being used for control.
Measurement of a qubit collapses its corresponding area in the space vector, lessening the complexity of the state.
This process is similar to pooling in traditional CNN, so the controlled-$V$ gates perform the pooling layers in QCNN and the $U$ gates perform the convolution layers.
The connectivity inside QCNN reduces in a binary-tree fashion, which poses interesting compilation problems.
Although the first convolution layer can be easily implemented with nearest neighbor connections, non-nearest neighbor interactions are necessary in the subsequent convolution layers.
Note that classically controlled-$V$ gates do not need connection on the quantum architecture.

%% file: 4_architecture_space.tex
\section{Architecture Space}
\label{sec:architecture}

\begin{figure}
  \centering
  \includegraphics[width=0.95\linewidth]{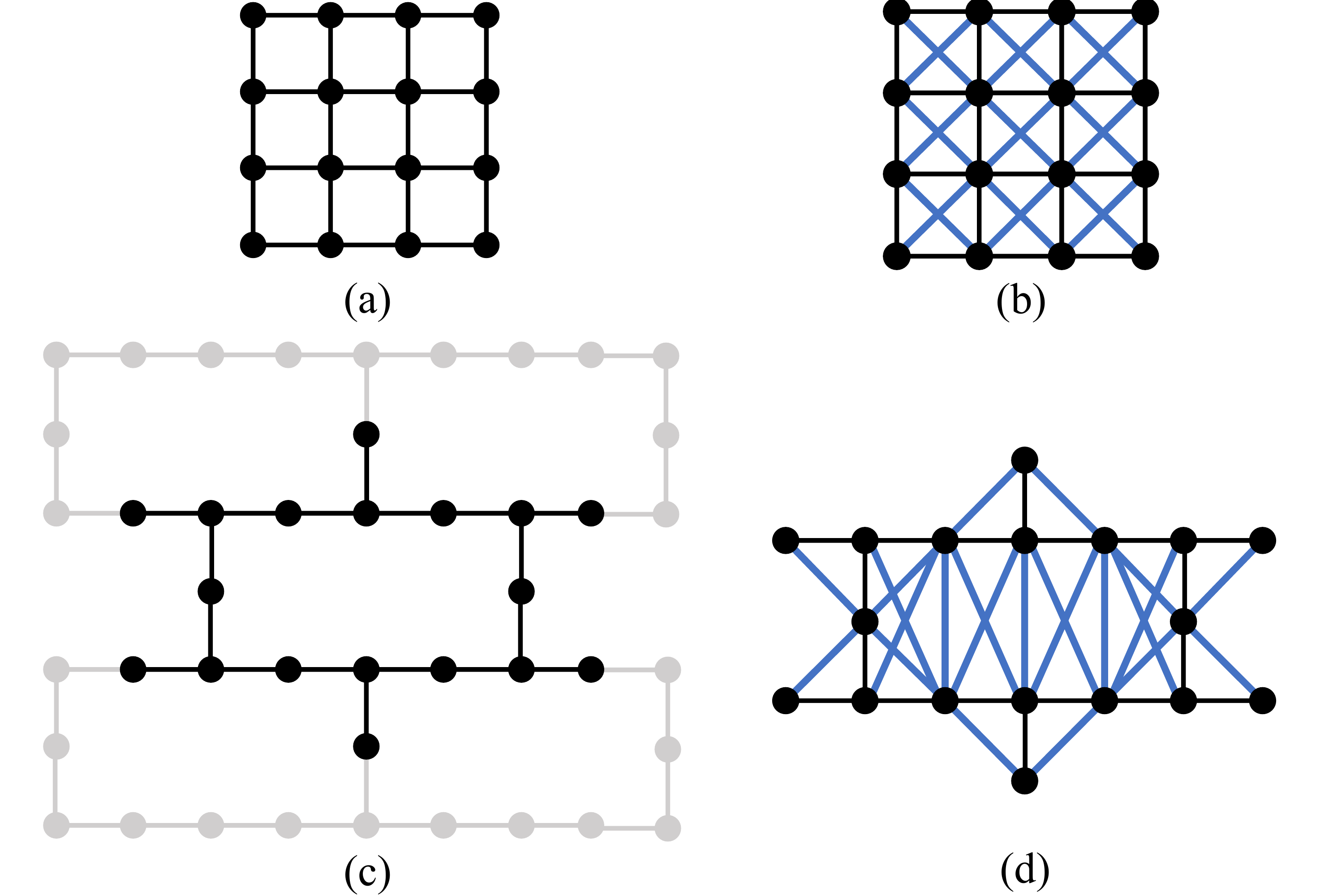}
  \caption{Architecture space. The black vertices and edges are physical qubits and fixed connections between physical qubits, respectively. The blue edges are the flexible qubit connections in our architecture space. The gray vertices and edges are parts of the heavy-hexagon architecture not included in the heavy-hexagon architecture space.}
  \label{fig:arch}
\end{figure}

Superconducting qubits are one of the most promising approaches towards scalable quantum computation.
Different superconducting quantum architectures support different qubit connectivities.
Qubit connectivity is defined by a coupling graph $G=(V,E)$.
This coupling graph has both a vertex $v \in V$ that denotes a physical qubit and two vertices $v, v'$ that only share an edge $e\in E$ if two-qubit entangling operations can be performed on physical qubits $v$ and $v'$.
For example, Fig.~\ref{fig:arch}a displays a coupling graph for a grid architecture with 16 physical qubits and 24 qubit connections, and Fig.~\ref{fig:arch}c shows a different coupling graph for a heavy-hexagon architecture with 18 physical qubits and 18 qubit connections.
Architecture customization can be performed during the fabrication or post fabrication.
In order to customizing architectures in post fabrication, one can construct programmable architectures.
For example, the qubit connections can be realized by tunable-coupler technology~\cite{nature19-google-quantum-supremacy}, which can turn off the superconducting qubit coupling via frequency adjustment, and thus, change the qubit connectivity.

The quantum architecture space can be defined by a triple $(G_b, E_f, C)$, where $G_b=(V,E)$ is a base coupling graph with fixed edges.
The flexible edge set $E_f$ is the set of edges that can be turned on, i.e., \emph{activated}.
Because activating some edges simultaneously may be impractical due to physical and fabrication constraints, $C$ is the collision edge pair set such that if $(e,e')\in C$, then edge $e$ and $e'$ cannot both be on at the same time. 
For example, we are not allowed to enable edges that cross the other edges due to the space capacity constraints between qubits in the existing quantum hardware.

In this work, we focus on two families of existing quantum architectures designed for scalable quantum error correction code: the grid architecture (Fig.~\ref{fig:arch}a), which is similar to those used by Google~\cite{nature19-google-quantum-supremacy}, and the heavy-hexagon architecture (Fig.~\ref{fig:arch}c), which is similar to those used by IBM~\cite{blog2021_IBM_hex}.
All vertices and edges in Fig.~\ref{fig:arch}c form a recurrent heavy-hexagon architecture.
The tiled layout of the heavy-hexagon architecture is chosen to preserve the heavy-hexagon structure and have the comparable physical qubit number to the grid architecture.
The black edges in Fig.~\ref{fig:arch}c represent the base heavy-hexagon architecture, also called fixed edges. 
These edges stand for couplings between qubits that are fundamentally fixed and cannot be modified. 
This constraint can come from practical concerns or limitations in quantum system engineering.
Note that four degree-one vertices are included to preserve degree-three vertices in the structure so that the degree-three vertices can entangle with three different qubits.

The grid architecture search space is shown in Fig.~\ref{fig:arch}b, where the base coupling graph is denoted in black and the flexible edge set is indicated in blue. 
A flexible edge is constructed between two vertices if their Euclidean distance is two and they are in different columns and rows.
Therefore, in the case we study here, the flexible edge set is all diagonal edges.
Due to fabrication concerns, all crossing edge pairs are added to the collision edge pair set $C$.

The heavy-hexagon architecture search space is depicted in Fig.~\ref{fig:arch}d.
The flexible edge set is constructed as follows.
First, we build the flexible vertical edges between non-adjacent vertices in the same column.
Note that there are no edges between two degree-one vertices because both vertices are outside the current rectangle.
Then, we construct the flexible edges between two vertices with distance two and in different columns and rows. 

Flexible edges in a coupling graph can be implemented by injecting additional quantum control to enable or disable a coupling. 
This is realizable in both superconducting qubits, and in ion-trap qubit systems. The versatility provided by the flexible edge architecture increases the range of applicable domains for a given quantum architecture. 
However, adding flexible edges can also induce higher overhead in the amount of hardware engineering and cost. Therefore, the architecture optimization framework proposed in this work will account  for the flexible edge and the fixed coupling edge differently to demonstrate the advantages that can come with a flexible coupling edge.

%% file: 5_search_strategy.tex
\section{Quantum Architecture Optimization}
\label{sec:qas}

Our domain-specific quantum architecture optimization framework facilitates the performance improvement of NISQ algorithms by adapting the quantum architecture to the quantum circuits of the specific application. Our proposal integrates the optimal layout synthesis into the subroutine of our architecture optimization to guarantee efficiency, scalability and performance.   

To perform architecture optimization, we need to address two challenges, flexible edge selection and large architecture space.
First, the optimality for the searched architecture is affected by the flexible edge selection.
Although activating more flexible edges will decrease the compilation overhead, it could introduce more crosstalk errors in to the circuit, which might eventually eliminate the benefit of activating edges.
For hardware design, minimizing the number of activate flexible edges is favorable; otherwise, it will be difficult to fabricate and calibrate.
In addition, deciding which set of flexible edges to activate are important.
Activating flexible edges at different locations brings varied levels of improvement.
Furthermore, not all flexible edges can be activated at the same time due to the fabrication limit.
Therefore, it is crucial to select the set of flexible edges that brings the most benefits to activate.

To guarantee the optimality, one naive approach for architecture optimization is to exhaustively evaluate all architectures in the search space using a layout synthesis tool and to select the architecture with the highest estimated circuit fidelity. 
However, this approach is inefficient since the size of the search space is exponential in that of the flexible edge set, e.g., $2^{18}$ for the grid architecture in Fig.~\ref{fig:arch}b.
On top of this, architecture evaluation requires a  near-optimal or optimal layout synthesis tool.
Otherwise, sub-optimal compilation would falsely favor some architectures, leading to bias in the evaluation results.
Previous implementations for optimal SWAP insertion have been proposed, yet these approaches suffer from very high space and time complexity~\cite{dac19-wille-burgholzer-zulehner-mapping-minimal-swaph,aspdac14-wille-lye-drechsler-optimal-swap-insertion,cgo18-siraichi-santos-collange-pereira-qubit-allocation}.
Some heuristic approaches have also been developed~\cite{web18-ho-bacon-cirq,web18-ibm-qiskit,date18-zulehner-paler-wille-efficient-mapping-ibmqx,qst20-sivarajah-dikes-cowtan-simmons-edgington-duncan-tket-compiler-nisq,asplos19-li-ding-xie-sabre-mapping,isca19-murali-linke-martonisi-abhari-nguyen-alderete-triq-architecture-studies,tcad08-maslov-falconer-mosca-placement, murali2019formal_constraint_based_compilation_for_nisq}, but have been shown to be far from optimal due to early termination in search trees~\cite{tc20-tan-cong-optimality-layout-queko}.

To address the aforementioned issues, we propose a novel quantum architecture optimization algorithm that integrates the optimal layout synthesis process into the architecture optimization procedure.
The key idea is to make the optimal layout synthesizer automatically explore the architecture with optimal synthesis results. 
We extend the layout synthesizer to make it capable to use a limited number of flexible edges during the compilation.
In this way, the layout synthesizer can automatically select the most beneficial flexible edges for the compilation, and the architecture optimizer can recognize the beneficial edges according to the layout synthesizer's choice.
Therefore, we can avoid enumerating and evaluating all architectures individually, and instead, our integrated algorithm minimizes the number of layout synthesis invocations due to a more efficient optimization.
For example, a quantum circuit of which compilation takes one hour on the grid architecture in Fig~\ref{fig:arch}b in one hour will require $2^{18}$ hours to explore the whole design space. 
This assumes no additional compilation cost per flexible edge.
However, according to our experiments, our algorithm only needs a few hours to finish the optimization process.

The recent layout synthesis tool OLSQ \cite{iccad20-tan-cong-optimal-layout-synthesis} is optimal in terms of SWAP count and depth by formulating the layout synthesis problem as an satisfiability modulo theory (SMT) \cite{barrett2018-smt} problem.
The SMT problem is a generalization of the Boolean satisfiability problem by including the other first-order theories e.g., equality reasoning and arithmetic. 
With its strong expressive power, a wide range of applications, e.g., bounded model checking, program analysis, and software testing, can be formulated as SMT problems.
A more coarse-grained, transition-based implementation, referred to as TB-OLSQ, is also developed to utilize an efﬁcient variable encoding to represent the layout synthesis solution space, including both qubit mapping and SWAP insertion, and preserves optimality in terms of SWAP insertion and near-optimal in terms of depth.
Furthermore, TB-OLSQ matches optimal results on verifiable instances while outperforming leading heuristic layout synthesis tools on gate count optimization. 
Since the main benefit of customizing qubit connectivity is to reduce the required number of inserted SWAP gates, achieving optimal SWAP insertion is one of the most important objectives.
With good scalability and optimal compilation results, TB-OLSQ is the choice of our layout synthesizer,
which provides efficient runtime, minimal SWAP count, and enables us to maximize circuit fidelity in a scalable manner.

\begin{figure}
  \centering
  \includegraphics[width=\linewidth]{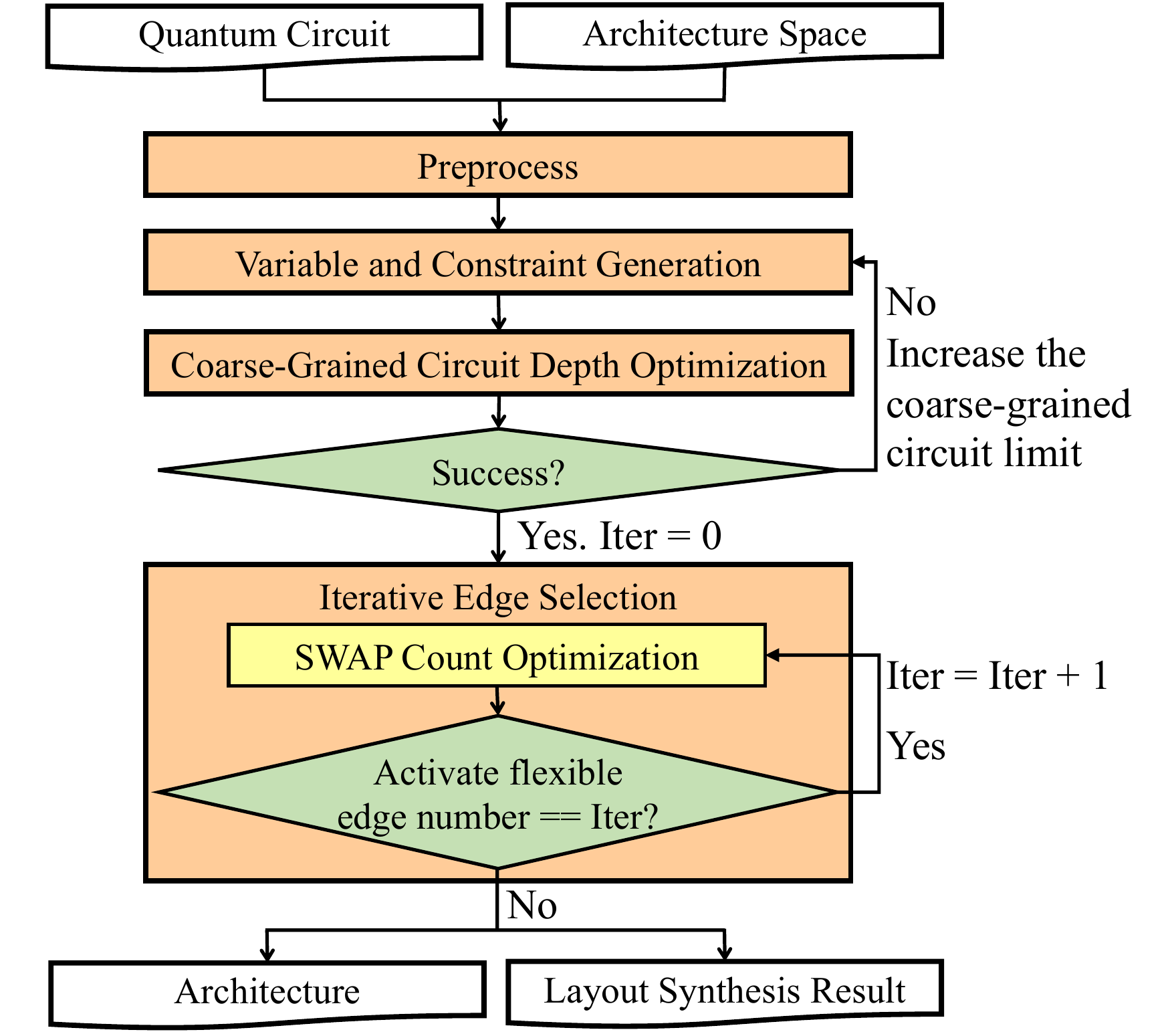}
  \caption{Overview of our quantum architecture optimization flow.}
  \label{fig:flow}
\end{figure}

Given a quantum circuit consisting of single- and two-qubit gates and an architecture space specified by the three-tuple $(G_b,E_f,C)$, 
our proposed algorithm outputs the optimal architecture for the circuit and the layout synthesis results.
The three major stages consist of: (1) SMT variable and constraint generation, (2) coarse-grained circuit moment optimization and (3) iterative edge selection accounting for SWAP count optimization.
Fig.~\ref{fig:flow} illustrates the overall workflow. 
The following subsections describe each stage in detail.

\subsection{Preprocessing}
In order to generate variables and constraints for the flexible edges in the search space, we first construct a gate dependency list $L$ from the input circuit to impose temporal constraints such that gates acting on the same qubit cannot be executed during the same circuit moment.
Thus, the gate dependency list is constructed by adding all gate pairs $(g,\ g')$ to $L$, where $g,\ g'$ act on the same logical qubit and $g$ should be executed before $g'$. 
With the gate dependency list, we can then generate constraints that enforce the correct gate execution order.

\subsection{Variable and Constraint Generation}
Following propocessing, we generate variables and constraints to formulate our problem as an SMT instance.
We utilize circuit-related variables and constraints conforming to the TB-OLSQ formulation to encode the layout synthesis problem, and architecture-related variables and constraints to encode the architecture optimization problem.
We use $G_1$ to denote the single-qubit gate set, $G_2$ to designate the two-qubit gate set, $v\in V$ for physical qubits, and $e\in E\cup E_f$ for qubit connections.
Additionally, we define a initial upper bound $T_{max}$ for the coarse-grained circuit depth.
Henceforth, for simplicity, circuit moments and depth will be assumed to be coarse-grained in this section unless otherwise specified.

\subsubsection{Variables}
We create circuit-related variables for qubit mapping, coarse-grained circuit moment index, space coordinate, and SWAP insertion.
For each logical qubit $q$ and coarse-grained circuit moment $t$, we define the map variable $\pi^t_q$ as the mapping of the logical qubit $q$ to the physical qubit $v$ at the coarse-grained circuit moment $t$.
Note that the circuit depth increases when the qubit mapping is changed, i.e., a qubit mapping \emph{transition} occurs, since a SWAP operation is inserted.
Therefore, the coarse-grained circuit depth is the number of total transitions.
Accordingly, gates acting on the same qubits can be executed within the same coarse-grained circuit moment, i.e., in between two adjacent transitions.
Each gate $g$ has a space coordinate $x_g$ and moment $t_g$ indicating that gate $g$ is scheduled on $x_g$ at moment $t_g$.
If $g\in G_1$, then $x_g$ is some physical qubit $v\in V$.
If $g\in G_2$, then $x_g$ is some edge $e\in E\cup E_f$.
The boolean SWAP insertion variable $\sigma^t_k$ represents the application of a SWAP gate on edge $e_k$ at moment $t$, if $\sigma^t_k=1$.
Then, we construct architecture-related variables for the flexible edge activation.
Each flexible edge $e\in E_f$ has a boolean variable $u_{e}$, where $u_e = 1$ if at least one gate is scheduled to be executed on $e$.

\subsubsection{Constraints}
Program constraints consists of mapping, dependency, SWAP insertion, and SWAP transformation constraints.
Architecture constraints consist of flexible edge usage and edge validity.

\noindent \textbf{Mapping.}
For map variables in the same moment $0\leq t \leq T_{max}$, we require $\pi^t_q \neq \pi^{t}_{q'}$ so that logical qubits $q \neq q'$ are mapped to distinct physical qubits.
Also, the qubit mapping must be consistent with the gate mapping. 
For a gate $g\in G_1$ acting on physical qubit $v$ and circuit moment $t$, we have:
\begin{equation}
    (x_g == v) \wedge (t_g == t)\implies (\pi^t_q==v).
\end{equation}
For a gate $g\in G_2$ acting on edge $e=(v_1,v_2)\in E\cup E_f$, we have:
\begin{align}
    \nonumber &(x_g == e) \wedge (t_g == t)\implies \\ &(\pi^t_q==v_1\wedge\pi^t_{q'}==v_2)\vee(\pi^t_q==v_2\wedge\pi^t_{q'}==v_1).
\end{align}

\noindent \textbf{Dependency.}
To maintain the gate execution order in the input circuit and prevent collisions, we have $t_g \leq t_{g'}$ for each pair $(g,g')\in L$.

\noindent \textbf{SWAP Insertion.} To avoid two SWAP gates $\sigma_e^t$, $\sigma_{e'}^{t'}$ from overlapping a physical qubit, we have $\neg\sigma_e^t \vee \neg\sigma_{e'}^t$ so that these two SWAP gates are in different circuit moments.

\noindent \textbf{SWAP Transformation.}
A SWAP insertion requires two additional constraints on the qubit mapping.
First, the mapping for the logical qubit $q$ to the physical qubit $v$ should not change if no SWAP gate is inserted on any edge $e\in S_v=\{(v,v')|v'\in V\}$ in moment $t$:
\begin{equation}
    [(\pi^t_q == v)\wedge \bigwedge_{e\in S_v}\neg\sigma^t_e]\implies (\pi^{t+1}_q == v).
\end{equation}
Second, the mapping for the logical qubit $q$ to the physical qubit $v$ will change if a SWAP gate is inserted on one edge $e\in S_v$ in moment $t$:
\begin{equation}
    (\pi^t_q == v)\wedge \sigma^t_e \implies (\pi^{t+1}_q == v').
\end{equation}

\noindent \textbf{Flexible edge usage.} If there is one two-qubit gate $g\in G_2$ or a SWAP gate acting on edge $e\in E_f$, then $u_{e}=1$:
\begin{equation}
    \bigvee_{g\in G_2} (x_g == e) \vee \bigvee_{0 \leq t \leq T_{max}}\sigma^t_{e}\implies  u_{e}.
\end{equation}

\noindent \textbf{Edge validity.} For each edge pair $(e,e')$ in the collision edge pair set $C$, we have $\neg(u_{e}\wedge u_{e'})$ so that edge $e$ and $e'$ are not added to an architecture at the same time.

\subsection{Coarse-grained Circuit Depth Optimization}
\label{subsec:qas:coarse_circuit_depth_opt}

The purpose of this stage is mainly to find the minimum coarse-grained circuit depth for the base architecture, and in doing so, we also detect whether or not the upper bound $T_{max}$ is sufficiently large.
Since the activation of flexible edges in $E_f$ increases the connectivity of the coupling graph, the number of essential SWAP insertions could decrease as a result.
For this stage, however, we are only concerned with establishing a baseline minimum depth without activating any edges, which can be described with the constraint:

\begin{equation}
\sum_{e\in E_f}u_e\leq \alpha,
\label{eq:e_use_constraints}
\end{equation}
where $\alpha$ is the number of activated flexible edges and set to 0 for the baseline.
Note that Eq.~\ref{eq:e_use_constraints} can be extended to the weighted sum over $u_e$ if the cost of activating distinct flexible edges is different.
The coarse-grained circuit depth optimization can be performed via the constraint:
\begin{equation}
\bigwedge_{t_g} (t_g \leq T),
\end{equation}
where $T$ is the current circuit depth in consideration.
We minimize this depth by using a linear search strategy.
We initially set $T=0$.
If no feasible solution exists for the current value of $T$, we increment by one and check the satisfiability again until we find a solution.
If we fail to find a feasible solution for all $T \leq T_{max}$, we increase $T_{max}$ and regenerate variables and constraints.
Similarly to TB-OLSQ, we use the Z3 SMT solver \cite{tacas08-demoura-bjorner-z3-smt-solver} to solve our problem.
Z3 provides us with two modes of execution: (1) check satisfiability only, and (2) check satisfiability and minimize an objective.
TB-OLSQ uses mode 2, but we found the increase walltime cost of this mode to be too expensive for our larger problems, so we chose to implement the previously described optimization routine around mode 1.

\subsection{Iterative Edge Selection}
\label{subsection:qas:edge_selection}
During iterative edge selection, we activate a bounded number of flexible edges while searching for the minimum number of SWAP insertions per iteration.
To achieve this, we set $\alpha$ from Eq.~\ref{eq:e_use_constraints} to $i$, where $i$ is the current iteration index, and then we perform the optimization by applying a constraint using the SWAP count limit $S$:
\begin{equation}
\sum_{0 \leq t \leq T, e\in E\cup E_f}\sigma^t_e\leq S.
\end{equation}
We employ binary search to minimize $S$.
The lower bound $S_l$ for $S$ is zero-initialized while the upper bound $S_u$ is initialized to the number of SWAP insertions from the previous stage.
Note that activating a flexible edge will never increase the optimal SWAP count due to the increased qubit connectivity.
Then, for each iteration $i$, if a solution exists at the midpoint $S = \lfloor (S_l+S_u)/2 \rfloor$, then the upper bound $S_u$ is updated to $S$. Otherwise, we update the lower bound $S_l$ to $S+1$.
The process repeats until $S_l \geq S_u$.
Then, we generate the layout synthesis solution under $S = S_u$.
The optimization process ends if the number of activated flexible edges in the architecture is less than $i$ or no SWAP gate is inserted.
Both cases indicate that activating more edges will not further benefit the circuit fidelity.
Since each iteration requires an invocation of the SMT solver, we reduce the amount of duplicated work by utilizing incremental solving, in which the previous effort from base layout synthesis is used as the starting point for each iteration.
Finally, we output the set of optimized architectures using a different number of activated flexible edges.

\subsection{Complexity and Scalability}
Our formulation contains $QT+2|G_1\cup G_2|+|E\cup E_f|T_{\mathit{max}}+|E_f|$ variables, where $Q$ is the number of logical qubits in a circuit.
Consider the fabrication limitations, the number of edges is usually asymptotically linear to the number of vertices.
Thus, the variable number in our formulation is $O(|V|\cdot T_{\mathit{max}}+|G_1\cup G_2|)$.
Note that the search space size is exponential in the number of variables.
Compared with the variable number of TB-OLSQ~\cite{iccad20-tan-cong-optimal-layout-synthesis}, introducing architecture-related variables do not increase the problem complexity.
With the incremental solving strategy, we reduce the total solving time by preventing variable and constraint regeneration and keeping the information learnt from the previous solving process in the later solving process.

%% file: 6_evaluation_metrics.tex
\section{Evaluation Metrics}
\label{sec:eval_m}
\begin{figure}
  \centering
  \includegraphics[width=\linewidth]{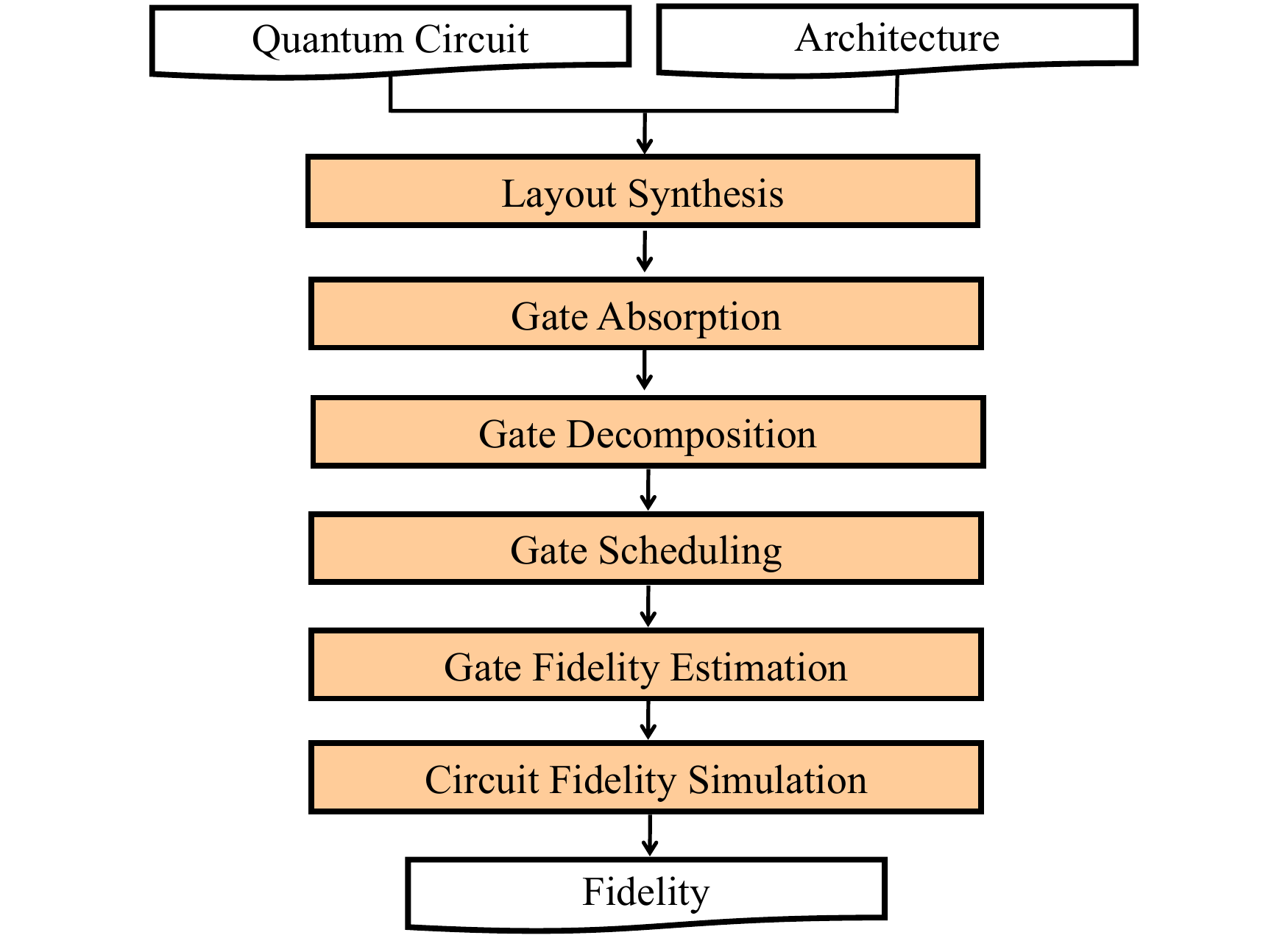}
  \caption{Overview of our evaluation flow.}
  \label{fig:eval_flow}
\end{figure}

In this section, we introduce our evaluation method for estimating the circuit performance of an architecture.
Fig.~\ref{fig:eval_flow} shows our evaluation workflow.
Given a quantum circuit and an architecture, we first perform layout synthesis to map the circuit to the architecture.
Then, when applicable, we apply gate absorption, i.e. merging consecutive two-qubit gates, to further optimize the circuit.
In our study, there are four types of gate pairs that can be merged: (1) two phase-splitting operators, (2) one phase-splitting operator and one SWAP gate, (3) two U4 gates, and (4) U4 gate and one SWAP gate.
In our case, the most frequent type is merging a SWAP gate with some other gates.
Therefore, the compilation results that contain more SWAP gates are more likely to be improved by gate absorption.
For example, in the QAOA-8 compilation results by OLSQ on the base heavy-hexagon architectures, up to six gates can be absorbed.
On average, we can absorb two gates in each compilation result.

Next, gates are decomposed to the hardware-supported gates and scheduled according to gate duration.
In this work, we use the Google Sycamore gate along with arbitrary single-qubit gates as our hardware-supported gate set~\cite{nature19-google-quantum-supremacy}.
The gate duration is 10$\mathrm{ns}$ for the Sycamore gate, $25\mathrm{ns}$ for single-qubit gates, and $4\mathrm{\mu s}$ for measurements~\cite{web21-google-sycamore-data-sheet}.
We schedule gates into different circuit moments such that all gates within the same moment can be executed simultaneously.
Note that for gates with short duration, the affected qubit will remain idle until the next circuit moment.

Based on the gate scheduling, we can calculate the two-qubit gate fidelity accounting for crosstalk error caused by parallel gate execution and estimate the circuit fidelity using our fidelity model. 
In the following sections, we will define our crosstalk metric and circuit fidelity model. Notice that the crosstalk error included here is a type of constraints for the underlying architecture optimization: each qubit cannot have too many neighbors; 
otherwise, practical limitations in classical control electronics will pose limitations on the achievable fidelity of quantum operations. 
In addition, there are potentially other constraints important to accommodate during qubit architecture optimization. 
These additional constraints may be derived from fundamental limitations of the underlying physical system. 
For example, for ion trap qubits, the maximum number of fully connected qubits can be limited, and for nitrogen-vacancy qubits, the available type of coupling might not be limited; etc. 
The advantage of our framework lies in the flexibility of SMT solvers for incorporating different types of hard constraints during optimization.

\subsection{Crosstalk Effect}
\label{sec:eval_m:crosstalk}
Unwanted couplings between qubits introduce crosstalk errors. 
The effect of the crosstalk error manifests as a decrease in circuit fidelity when quantum gates are operating in parallel as opposed to in isolation. 
Crosstalk in most quantum computing systems is electromagnetic in nature and falls into two categories: classical crosstalk and quantum crosstalk.
The effect of classical crosstalk causes independent and identically distributed (i.i.d.) errors across space due to shared electromagnetic fields.
This happens at the control-level electronics.
For example, two control wires that are too close to each other can cause control leakage as a result of flux crosstalk. 
The effect of quantum crosstalk is more diverse because it can be caused by various types of electromagnetic interaction, e.g., capacitive coupling, inductive coupling, and other secondary, non-linear interactions in superconducting circuits. 
 
We pay special attention to the mitigation of quantum crosstalk errors due to two fundamental reasons. First, the effect of quantum crosstalk can be classically hard to simulate given its complex form and propagation dynamics. Second, majority of the near-term  quantum algorithms and long-term quantum error correction algorithms are much more susceptible to crosstalk errors than to   i.i.d gate errors~\cite{dennis2002topological,kattemolle2022effects_of_correlated_errors_on_the_qaoa}.
To illustrate the detrimental effect of crosstalk, we take   quantum error correction as an example to show that the tolerence against crosstalk can be much lower than independent errors. This simple analysis is applicable to other NISQ algorithms sharing similar circuit structures. In quantum error correction, the tolerable level of two-body crosstalk error probability is determined by comparing correlated error probability from crosstalk with error coincidence from i.i.d gate errors:~\cite{dennis2002topological} \begin{equation}\label{eq:crosstalkthreshold}
     P_{\text{crosstalk}} <  P_{\text{ave}}^2,
\end{equation} 
where $P_{\text{ave}}$ is the correctable average i.i.d gate error. 
If this inequality is satisfied, two-body crosstalk errors become visually indistinguishable from i.i.d errors in its lower moments of distribution. 
Such requirement derives from the assumptions  in fault-tolerant computation that error propagate in a probabilistic manner and will not accumulate and amplify quantum-mechanically~\cite{book10-nielsen-chuang-quantum-computation-information}.  Inequality ~(\ref{eq:crosstalkthreshold}) shows that we have to suppress crosstalk error much harder than single gate errors to ensure that errors are within simulability and control.
For NISQ applications, the threshold above $P_{\text{ave}}$ can change depending on specific application requirements based on crosstalk error simulation.

 
Despite the distinct and diverse mechanisms of crosstalk, we can systematically measure the average crosstalk effect by comparing the fidelity of a quantum gate $g$ when operated in isolation $P_{g}$ with that of a maximally parallel configuration $P_{g}^p$, where we use superscript $p$ to denote parallel operation. 
The maximally parallel configuration is specified according to the hardware limitations, e.g., only quantum gates that do not share common qubits can be executed in parallel. 
For example, in the grid architecture, only one fourth of all two-qubit gates can be operated in parallel due to the degree four nature of the grid connectivity. 
Let the number of nearest neighboring gates in such a maximally parallel configuration be $N_1$, and the number of the neighbor gates of distance $i$ away be $N_i$, where $i\leq n_{max}$, which is a cutoff threshold beyond which the crosstalk induced error is below the threshold defined in Eq.~\ref{eq:crosstalkthreshold}.
The exact value of $n_{max}$ depends heavily on the specific system of interest.
For superconducting qubits, we set $n_{max}=2$ based on the quadratic decay of charge-charge interaction, which is one of the leading sources of long-range crosstalk in superconducting circuits.
 
To account for the distance-dependent nature of crosstalk, we weigh the contribution of a neighbor with distance $i$ by a factor of $r_i\leq1$.
We can therefore define the crosstalk error with the nearest neighbor gate as:
\begin{align}
P_{g_{ct}}^1= \frac{P_{g}^p - P_{g}}{\sum_{i=1}^{ n_{max}}r_i N_i}.
\end{align} 
The crosstalk error from the $i$th nearest neighboring gate can be calculated by multiplying the decay factor: $P^i_{g_{ct}}=P^1_{g_{ct}} r_i$.
In this work, we use $r_i=\frac{1}{10^{i-1}}$, $P_{g} = 0.9\%$, and $P_{g_{ct}}^1 = 0.5\%$~\cite{nature19-google-quantum-supremacy}.

According to the above model, the fidelity of a two-qubit gate $g$ in a given circuit moment is then given by:
\begin{align}
f_g = 1 - P_g - \sum_{g'} P_{g_{ct}}^1 - \sum_{g''} P_{g_{ct}}^2,
\end{align} 
where $g'$ and $g''$ are the respective two-qubit gates of distance one and two to $g$ in the given circuit moment.
In principle, we can include crosstalk between both single-qubit and two-qubit gates. 
However, for the modeling of superconducting qubits, we can assume that executing single-qubit gates in parallel does not induce crosstalk error.

Given a coupling graph $G=(V,E)$ and a two-qubit gate schedule specifying space-time coordinates, Algorithm~\ref{alg:gate_fidelity} calculates the two-qubit gate fidelity. 
In lines~\ref{alg:line:d1}--\ref{alg:line:d2}, we construct the distance-one and distance-two lists $d_1.[e]$ and $d_2.[e]$ for each $e\in E$.
An edge $e'\in E$ is added to $d_1.[e]$ ($d_2.[e]$) if the shortest path from one vertex of $e'$ to one vertex of $e$ is one (two).
Then, for each circuit moment, we calculate the two-qubit gate fidelity by subtracting crosstalk errors from concurrent distance-one and distance-two gates (lines~\ref{alg:line:time_slot_begin}--\ref{alg:line:time_slot_end}), and then output the two-qubit gate fidelity list $L$.

\begin{algorithm}[t]
\small
\caption{\emph{GateFidelity($\mathit{TS}$,$G$)}}
\label{alg:gate_fidelity}
\begin{algorithmic}[1]
\REQUIRE
    Two-qubit gate schedule $\mathit{TS}$ and coupling graph $G$
\ENSURE
    Two-qubit gate fidelity list $L$
\STATE construct distance-one edge list $d_1.[e]$ for each $e\in E$ \label{alg:line:d1}
\STATE construct distance-two edge list $d_2.[e]$ for each $e\in E$ \label{alg:line:d2}
\STATE $L$ = []
\FOR {moment in $\mathit{TS}$} \label{alg:line:time_slot_begin}
    \FOR {$g$ in moment}
        \STATE gate\_fidelity = 1 - $P_{g}$
        \FOR {$g'$ in moment}
            \IF{$g'.\mathit{pos}\in d_1[g.\mathit{pos}]$} \label{alg:line:in_d1}
                \STATE gate\_fidelity = gate\_fidelity - $P_{g_{ct}}^1$
            \ELSIF{$g'.\mathit{pos}\in d_2[g.\mathit{pos}]$}
                \STATE gate\_fidelity = gate\_fidelity - $P_{g_{ct}}^2$
            \ENDIF
        \ENDFOR
        \STATE $L$.append(gate\_fidelity)
    \ENDFOR
\ENDFOR \label{alg:line:time_slot_end}
\RETURN $L$
\end{algorithmic}
\end{algorithm}

\subsection{Overall Circuit Fidelity Estimation}
For NISQ applications, circuit fidelity is an important metric because it measures the success rate of the probabilistic outcome.
Circuit fidelity is influenced by qubit decoherence and gate fidelity.
Our fidelity model for an entire circuit is:
\begin{equation}
    f=f_1^{g_1}\times \prod_{f_g\in L}f_{g}\times \prod_{q\in Q}(1-\frac{1}{3}(\frac{1}{T_1}+\frac{1}{T_\phi})T^q_{\mathit{idle}}), \label{eq:fidelity}
\end{equation}
where $f_1$ is the fidelity of single-qubit gate, 
$g_1$ is the number of single-qubit gates, 
and $f_{g}$ is the fidelity of each two-qubit gate including crosstalk error.
The last expression in Eq.~\ref{eq:fidelity} is to model idling qubit error derived from~\cite{nature19-google-quantum-supremacy}, where
$T_1$ is the thermal relaxation time, $T_{\phi}$ is the pure dephasing time, and $T^q_{\mathit{idle}}$ is the idling time for each qubit.
Note that for QAOA circuits, the qubit lifetimes are equal to the duration of the circuit, whereas for QCNN, each qubit lifetime ends at its time of measurement.
We use $f_1=99.9\%$, $T_1=15\mu s$, and $T_{\phi}=25 \mu s$ as existing technological factors \cite{nature19-google-quantum-supremacy}.

%% file: 7_evaluation_results.tex
\section{Evaluation Results}
\label{sec:eval}
In this section, we present the results from our architecture optimization framework for the following benchmarks: 
(1) QAOA \cite{arxiv1411-farhi-goldstone-gutmann-qaoa} phase-splitting operator for random 3-regular graphs of size 8 and 10 generated by \textit{networkx} version 2.4~\cite{hagberg2008networkx},
and (2) QCNN \cite{natphys19-cong-choi-lukin-quantum-convolutional-neural} of size 8.
First, we show the optimized architectures for each benchmark.
Next, we evaluate these architectures using the metrics described in Section~\ref{sec:eval_m}, and compare our performance against architectures from different layout synthesizers.
Then, we analyze the effects of our optimization for the heavy-hexagon and grid architectures.
Finally, we discuss the generalizability of our optimized architectures.

We implemented our proposed algorithm in Python 3.6, and used the Z3's Python API (v4.8.8.0)~\cite{tacas08-demoura-bjorner-z3-smt-solver} for the SMT optimization.
All experiments were conducted on a Linux machine with Intel Xeon E5-2680 CPUs at 2.40GHz and 64 GB RAM.

\subsection{Optimized Architectures}
\label{subsec:search_results}
\begin{figure}
\centering
    \subfloat[Architectures for QAOA-8 with $\alpha$ ranging from 1 to 7.\label{fig:hh_qaoa_8_arch}] {\includegraphics[width=\linewidth]{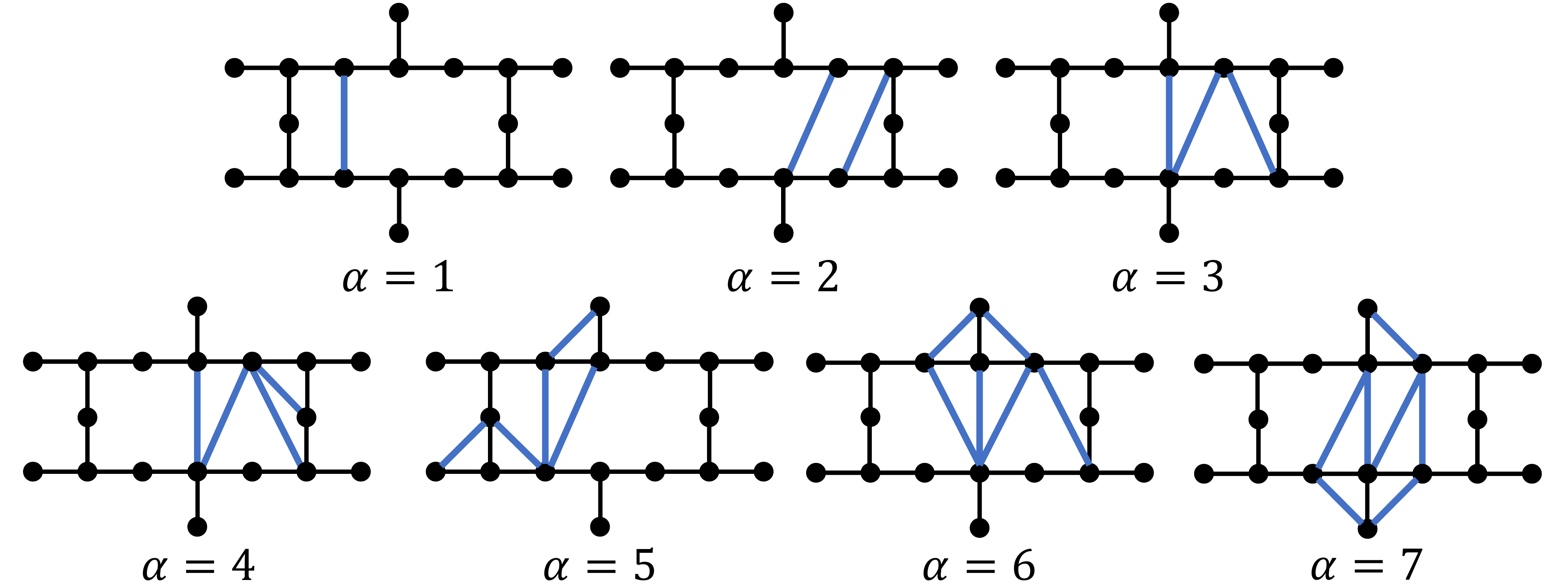}}
    
    \subfloat[Architectures for QAOA-10 with $\alpha$ ranging from 1 to 9.\label{fig:hh_qaoa_10_arch}] {\includegraphics[width=\linewidth]{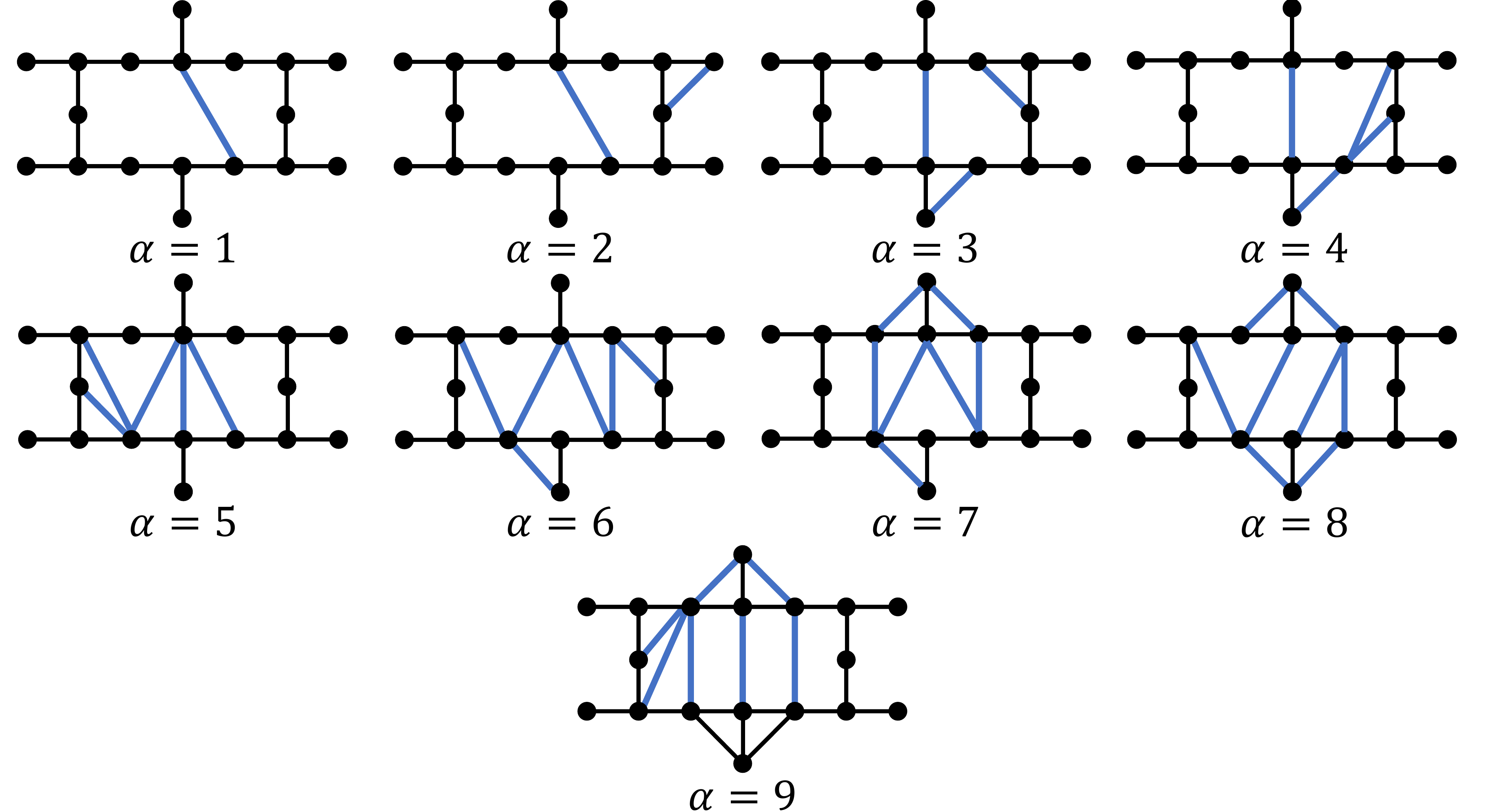}}
    
    \subfloat[Architectures for QCNN-8 with $\alpha$ ranging from 1 to 4.\label{fig:hh_qcnn_8_arch}] {\includegraphics[width=\linewidth]{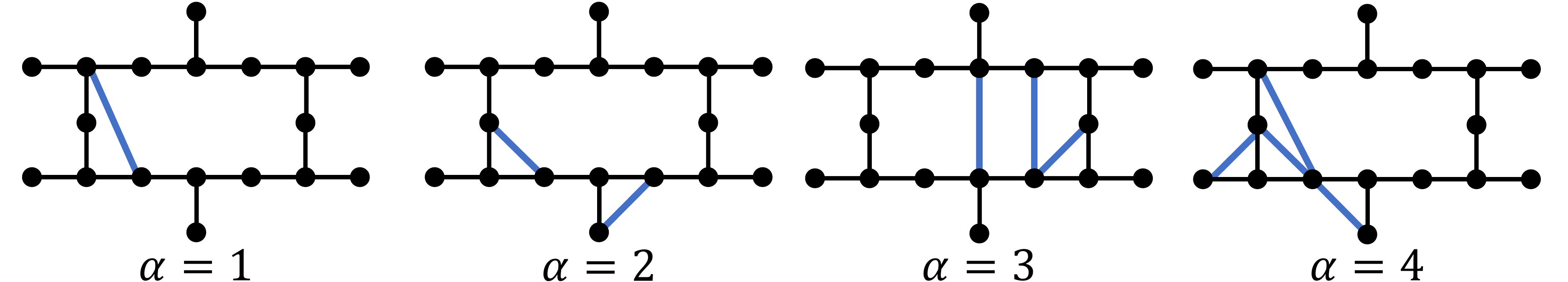}}

    \caption{Architectures optimized from heavy-hexagon architecture space, where the blue edges are the activated flexible edges.}
    \label{fig:hh_qaoa}
  \end{figure}

We show the architectures generated by our framework using the heavy-hexagon architecture space for QAOA and QCNN circuits in Fig.~\ref{fig:hh_qaoa}.
In general, the maximum value for the number of activated flexible edges increases as the number of essential SWAP insertion for compiling the circuit on the base architecture increases.
For example, QAOA-8 requires 8 SWAP gates and QAOA-10 requries 10 SWAP gates to execute on the base heavy-hexagon architecture, so the maximum number of activated flexible edges is 7 for QAOA-8 and 9 for QAOA-10 on the heavy-hexagon-based architectures.
The optimized heavy-hexagon architectures for QAOA-8 and QAOA-10 exhibit little inheritability, i.e. there are few common activated flexible edges across architectures, such as those in $\alpha=\{3, 4, 6\}$.
From this, we can observe that although some flexible edges are activated across different architectures, there is no common set of flexible edges that can be recognized as the most beneficial across all architectures.
Additionally, when comparing Fig.~\ref{fig:hh_qaoa_8_arch} and Fig.~\ref{fig:hh_qaoa_10_arch}, we do not observe any regularity for optimized architectures across different QAOA circuits for the heavy-hexagon architecture space.
This lack of regularity underscores the difficulty in discovering the optimal architecture through manual efforts and stresses the importance of having an effective and automated framework for architecture design.

\begin{figure}
\centering
    \subfloat[Architectures for QAOA-8 with $\alpha$ ranging from 1 to 3.\label{fig:grid_qaoa_8_arch}] {\includegraphics[width=\linewidth]{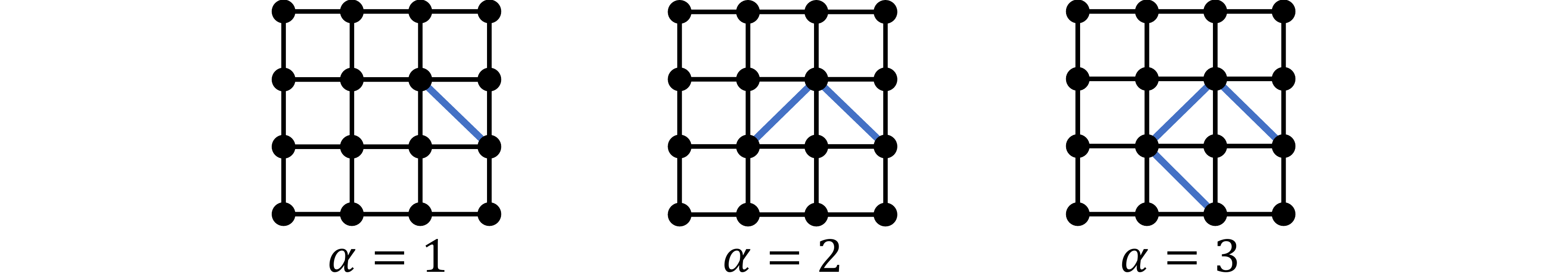}}
    
    \subfloat[Architectures for QAOA-10 with $\alpha$ ranging from 1 to 6.\label{fig:grid_qaoa_10_arch}] {\includegraphics[width=\linewidth]{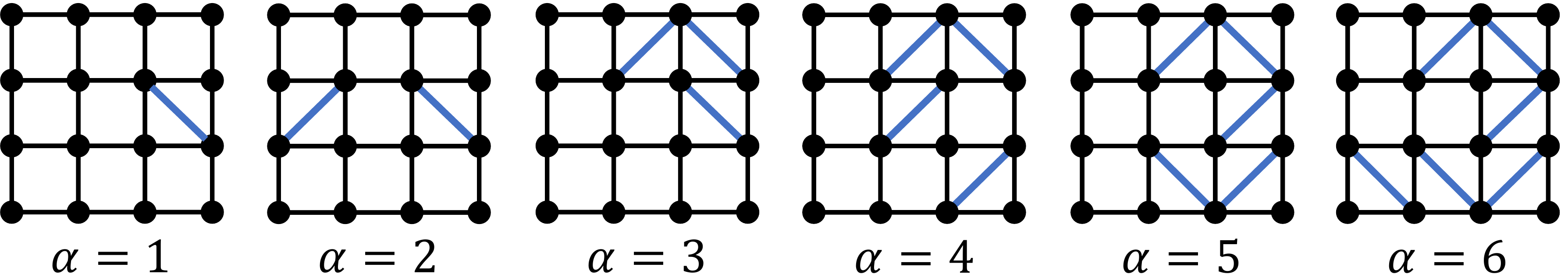}}
    
    \subfloat[Architectures for QCNN-8 with $\alpha$ ranging from 1 to 3.\label{fig:grid_qcnn_8_arch}] {\includegraphics[width=\linewidth]{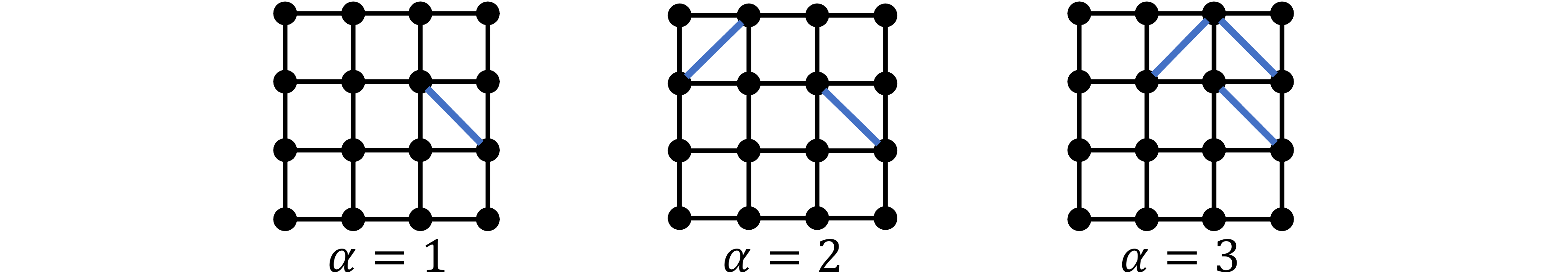}}
    \caption{Architectures optimized from grid architecture space where the blue edges are the activated flexible edges.}
    \label{fig:grid_qaoa}
  \end{figure}

The optimized grid-based architectures for QAOA and QCNN circuits are shown in Fig.~\ref{fig:grid_qaoa}.
Similarly to the heavy-hexagon study, more flexible edges are activated to reduce the need for SWAP gates as the circuit size increases.
Furthermore, the maximum value for $\alpha$ is smaller since there are inherently more fixed edges than heavy-hexagon-based architectures.
In contrast to the optimized heavy-hexagon-based architectures, the optimized grid-based architectures show high regularity, i.e. flexible edges activated in the previous $\alpha$ iteration tend to be activated again in the following.
Furthermore, the optimized architectures with $\alpha=3$ for QAOA-10 and QCNN-8 are the same, which suggests that the optimal grid-based architectures for different application instances may share similar structure patterns.
 
\subsection{Layout Synthesizer Comparison}
\label{subsec:compiler_comp}
To examine the fidelity improvement of architecture optimization, we utilize four layout synthesizers to map circuits to our optimized architectures. 
We chose two heuristic layout synthesizers, SABRE~\cite{asplos19-li-ding-xie-sabre-mapping} from Qiskit 0.20.1 and t$|$ket$\rangle$ 1.1.0~\cite{cowtan2019tketrouting}, and two optimal layout synthesizers, OLSQ and TB-OLSQ~\cite{iccad20-tan-cong-optimal-layout-synthesis} using Z3 version 4.8.8.0.

\subsubsection{QAOA}
\begin{figure}
\centering
    {\includegraphics[width=0.8\linewidth]{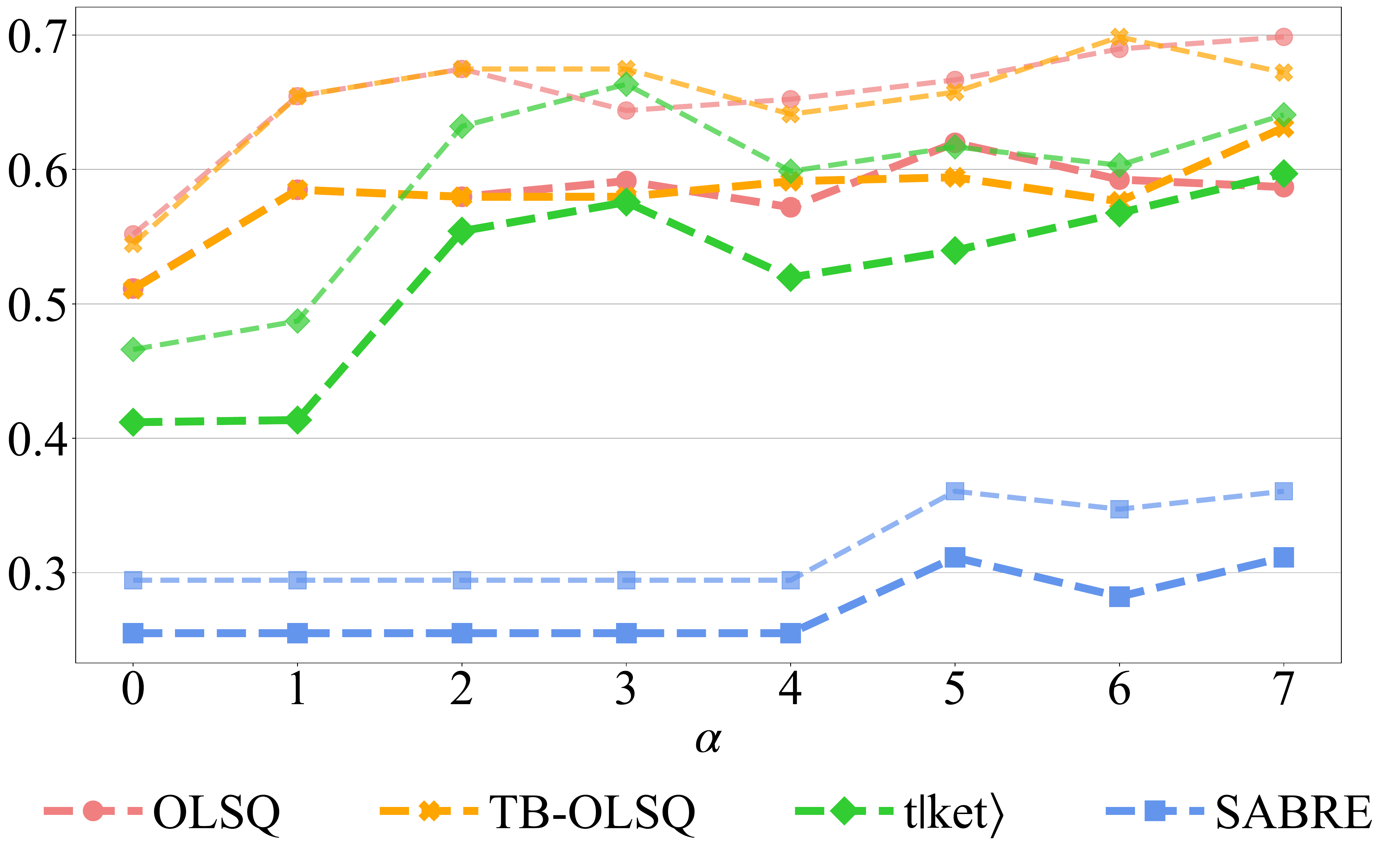}}
    

    \caption{Fidelity for QAOA-8 on the heavy-hexagon architecture space. Thick dashed lines are the fidelity estimates under crosstalk errors, and thin dashed lines are those without.\label{fig:c_hh_qaoa_8_fid}}
    \label{fig:c_hh_qaoa_8}
  \end{figure}

\begin{figure}
\centering
    {\includegraphics[width=0.8\linewidth]{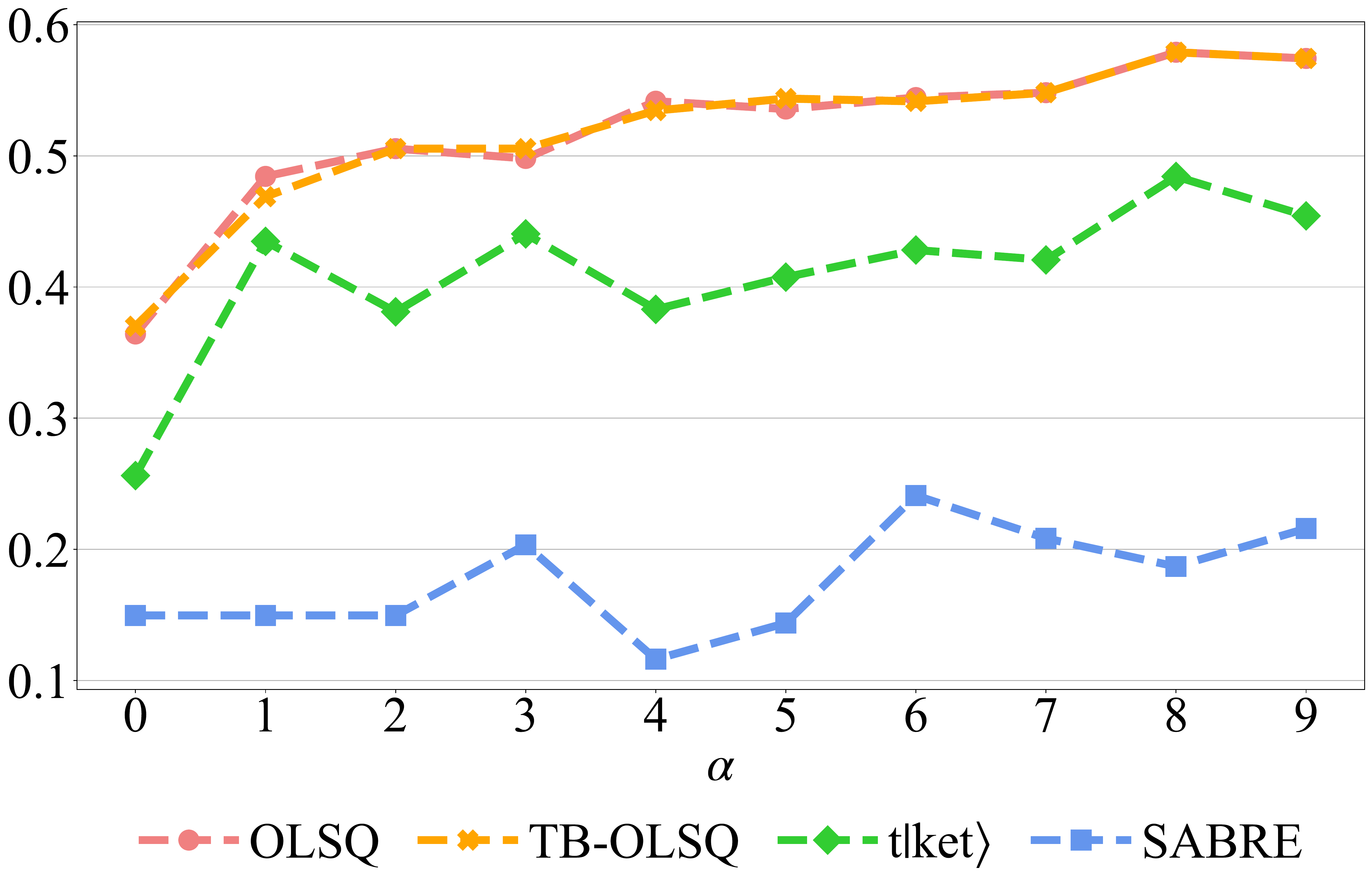}}
    

    \caption{Fidelity for QAOA-10 on the heavy-hexagon architecture space.}
    \label{fig:c_hh_qaoa_10}
  \end{figure}

The compilation results for QAOA-8 and QAOA-10 on the optimized heavy-hexagon-based architectures are shown in Fig.~\ref{fig:c_hh_qaoa_8} and Fig.~\ref{fig:c_hh_qaoa_10}, respectively.
We execute the circuit on the base architecture, i.e., $\alpha$ = 0, and use its fidelity as the baseline.
Note that activating more flexible edges may introduce additional crosstalk noises into the circuit, potentially eliminating benefits from increasing connectivity.
To analyze the factors causing fidelity changes, we calculate the circuit fidelity without crosstalk errors by assuming all fidelity values for two-qubit gates to be $1-P_g$.
In Fig.~\ref{fig:c_hh_qaoa_8_fid}, the thin dashed lines represent fidelity without crosstalk errors while the thick dashed lines represent fidelity with crosstalk errors.

In general, the optimized architectures achieve higher fidelity than the baseline (Fig.~\ref{fig:c_hh_qaoa_8}, Fig.~\ref{fig:c_hh_qaoa_10}).
Ignoring crosstalk effects, the circuit fidelity compiled by OLSQ and TB-OLSQ monotonically increases as flexible edges are activated.
This trend is in accordance with the assumption that activating more flexible edges improves fidelity by reducing the number of inserted SWAP gates.
However, the circuit fidelities compiled by SABRE do not show such a trend.
We posit that these underlying heuristic algorithms may fail to take advantage of the flexible edges, and thus fail to improve the circuit fidelity.
Among all layout synthesizers, SABRE produces the lowest fidelity and does not improve until five flexible edges are activated for QAOA-8 and three for QAOA-10.
In comparison, circuit fidelity compiled by the other layout synthesizers show improvement by only activating one flexible edge.
Based on this observation, we conclude that the advantage of architecture optimization would be limited if the solution quality of a layout synthesizer is unsatisfactory.

When accounting for crosstalk errors, the fidelity does not monotonically increase for all layout synthesizers.
As seen in Fig.~\ref{fig:c_hh_qaoa_8} and Fig.~\ref{fig:c_hh_qaoa_10}, some architectures with less activated flexible edges outperform those with more.
This decrease in fidelity for architectures with large $\alpha$ is due to crosstalk errors caused by the additional activated flexible edges.
This decline is more noticeable for OLSQ and TB-OLSQ since these optimal layout synthesizers tend to minimize the circuit depth by scheduling gates to be executed in parallel, resulting in more severe crosstalk errors.
For example, the architecture with $\alpha=4$ has relatively lower fidelity than the architectures with $\alpha=\{3,5\}$.
This is because the compilation result on the architectures with $\alpha=4$ has lower average two-qubit gate fidelity and longer qubit idling time than that of architectures with $\alpha=\{3, 5\}$
Therefore, for all figures, the fidelities for the optimized architectures do not exhibit a consistent increasing or decreasing trend due to this behavior.
Additionally, we note that through architecture optimization, the optimality gap between t$|$ket$\rangle$ and the optimal layout synthesizers shrinks for some optimized architectures (Fig.~\ref{fig:c_hh_qaoa_8}).

Fig.~\ref{fig:c_hh_qaoa_8} shows the fidelity improvement evaluated by OLSQ.
For QAOA-8, the fidelity peaks at the architecture with $\alpha=5$ and achieves a 21.1\% improvement compared to the baseline.
From Fig.~\ref{fig:c_hh_qaoa_10}, we observe a larger improvement for QAOA-10 than for QAOA-8, where fidelity improvement peaks at 59.0\% for the architecture with $\alpha=8$.

\begin{figure}
\centering
    {\includegraphics[width=0.8\linewidth]{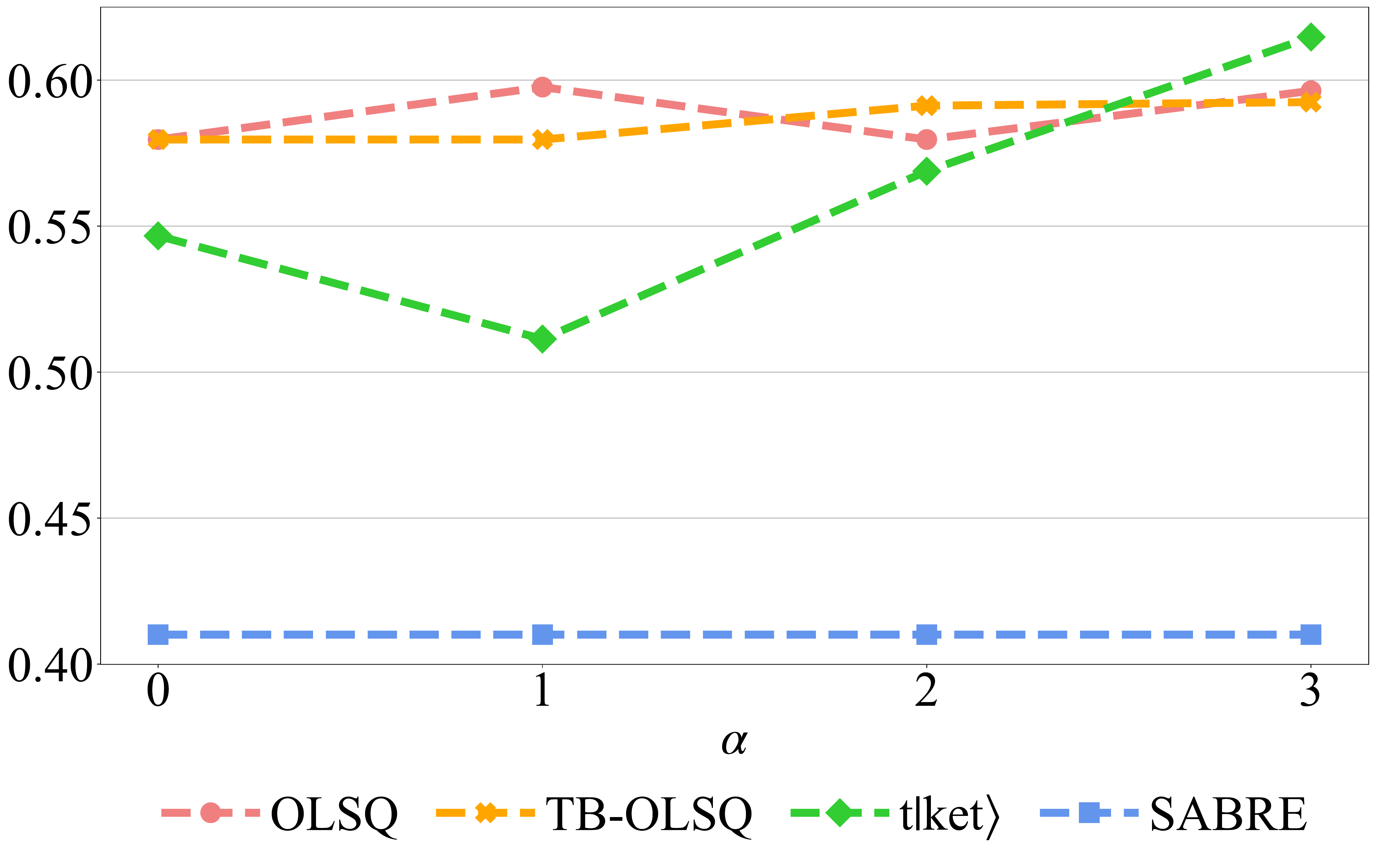}}
    

    \caption{Fidelity for QAOA-8 on the grid architecture space.}
    \label{fig:c_grid_qaoa_8}
  \end{figure}

\begin{figure}
\centering
    {\includegraphics[width=0.8\linewidth]{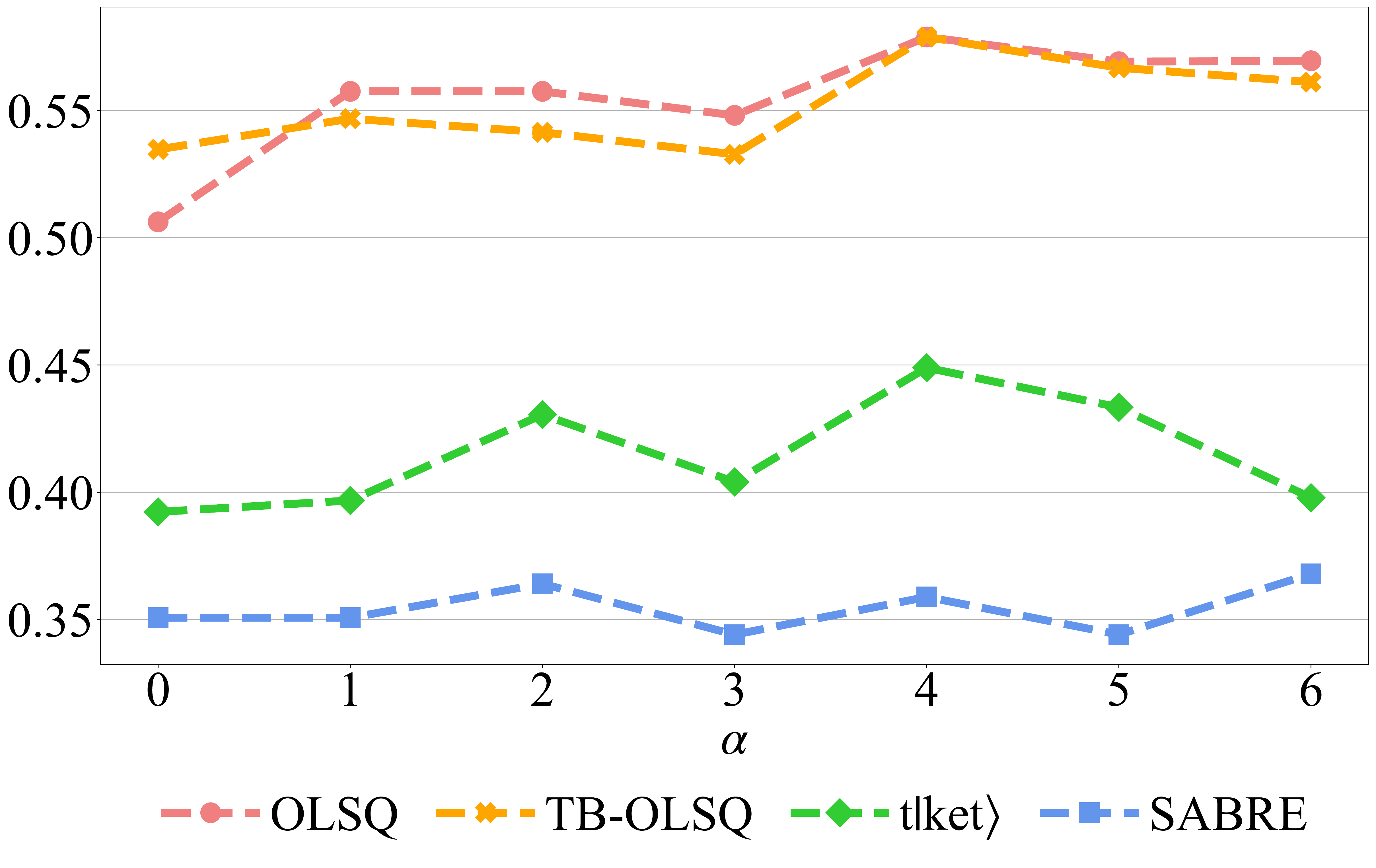}}
    

    \caption{Fidelity for QAOA-10 on the grid architecture space.}
    \label{fig:c_grid_qaoa_10}
  \end{figure}

Fig.~\ref{fig:c_grid_qaoa_8} and Fig.~\ref{fig:c_grid_qaoa_10} illustrate the results for QAOA compiled for the optimized grid-based architectures.
Similarly to before, Fig.~\ref{fig:c_grid_qaoa_8} and Fig.~\ref{fig:c_grid_qaoa_10} also shows that sub-optimal layout synthesizers cannot fully utilize the advantage of architecture optimization.
Likewise, the fidelity improvement from architecture optimization on grid-based architectures is also affected negatively by increases in crosstalk errors.
The optimal grid-based architectures for QAOA-8 and QAOA-10 are those with $\alpha=1$ and $\alpha=4$, which improve circuit fidelity by 3.1\% and 14.4\%, respectively 
(Fig.~\ref{fig:c_grid_qaoa_8}, Fig.~\ref{fig:c_grid_qaoa_10}).
Compared to the heavy-hexagon architecture space, optimization on grid-based architectures achieves less fidelity improvement in this case.
The first reason is that the connectivity of the base grid architecture can support the necessary connections for QAOA since these circuits require only three SWAP gates on the base grid architecture.
Second, the dense connectivity in grid-based architecture leads to more severe crosstalk errors, which can eliminate the benefit of architecture optimization.

\subsubsection{QCNN}
\begin{figure}
\centering
    {\includegraphics[width=0.8\linewidth]{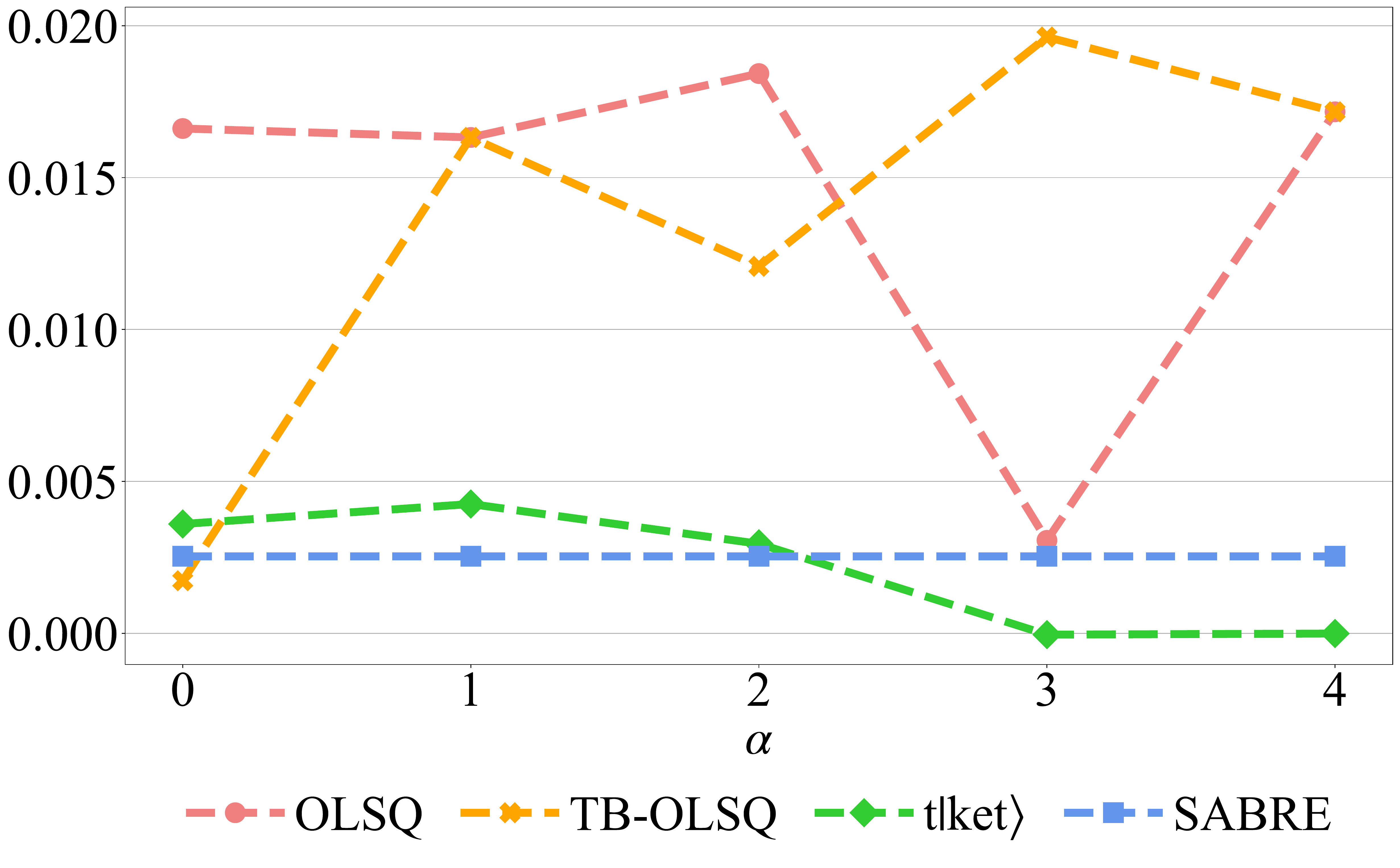}}
    

    \caption{Fidelity for QCNN-8 on the heavy-hexagon architecture space.}
    \label{fig:c_hh_qcnn_8}
  \end{figure}

\begin{figure}
\centering
    {\includegraphics[width=0.8\linewidth]{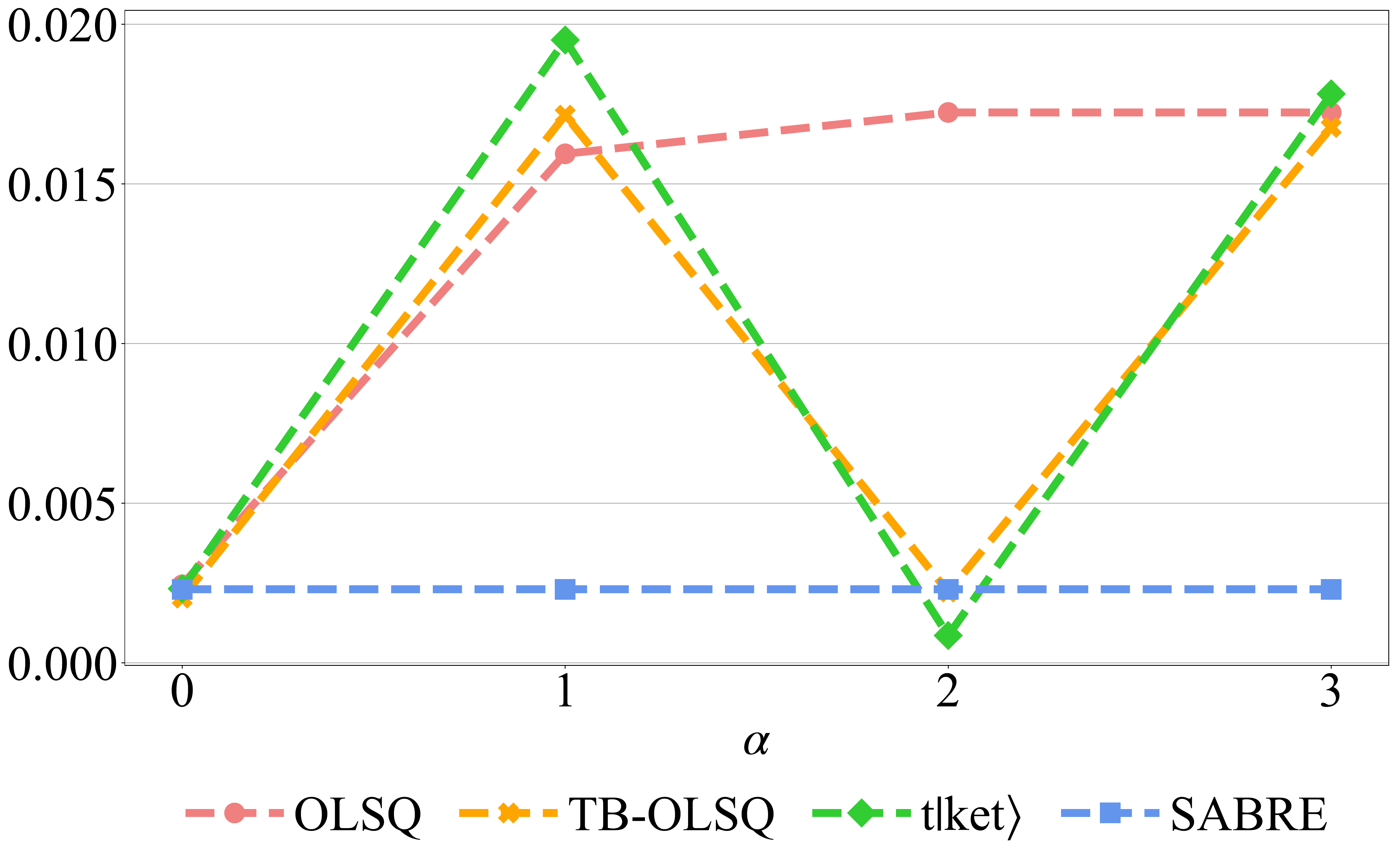}}
    

    \caption{Fidelity for QCNN-8 on the grid architecture space.}
    \label{fig:c_grid_qcnn_8}
  \end{figure}

We present the evaluation results for QCNN-8 in Fig.~\ref{fig:c_hh_qcnn_8} and Fig.~\ref{fig:c_grid_qcnn_8}.
Unlike the results from QAOA, we observe that an increase in crosstalk error due to the activation of more flexible edges does not greatly affect relative performance since QCNN-8 is dominated by qubit idling error from measurement-controlled gates.
With multiple pooling layers, qubits experience long idling times, which is ultimately detrimental to the circuit fidelity.

The peak performance for optimized heavy-hexagon-based architectures is 10.9\% with $\alpha=2$ (Fig.~\ref{fig:c_hh_qcnn_8}).
The fidelity for the heavy-hexagon-based architecture with $\alpha=3$ is lower than the others due to the long idling time $T^q_{\mathit{idle}}\approx 10.455\mu s$ while the idling time for the others is around $9\mu s$.
For the grid-based architectures, the architecture with $\alpha=2$, as seen in Fig.~\ref{fig:grid_qcnn_8_arch}, is the optimal architecture with a 604.8\% fidelity improvement.

Note that the duration of operations scheduled in the same circuit moment may be different. 
Those qubits whose operations are finished first will not proceed to the next operations until all operations scheduled in the same circuit moment are finished.
Nevertheless, when scheduling gates, the compilers assume duration for all gates to be the same.
Therefore, operations with different duration may be scheduled in the same circuit moment.
For QCNN-8, because the duration of measurement operations is much longer than that of the other operations, scheduling measurement operations in different circuit moments results in long idling time, which accounts for the performance gap between OLSQ and TB-OLSQ. Notice that such limitation is unique to superconducting qubit architecture under the consideration, and not inherent to the quantum application or our optimization algorithm.

\subsection{Architecture Space Comparison}
\label{subsec:arch_space_comp}
To study different architecture spaces, we compare the performance of optimized grid-based and heavy-hexagon-based architectures evaluated on QAOA and QCNN circuits.
We use OLSQ to perform layout synthesis since this produced the best solution quality in the previous experiments.

\subsubsection{QAOA}

\begin{figure}
\centering
    \subfloat[QAOA-8. Thick dashed lines are the fidelity estimates under crosstalk errors, and thin dashed lines are those without.\label{fig:qaoa_8}]
    {\includegraphics[width=0.8\linewidth]{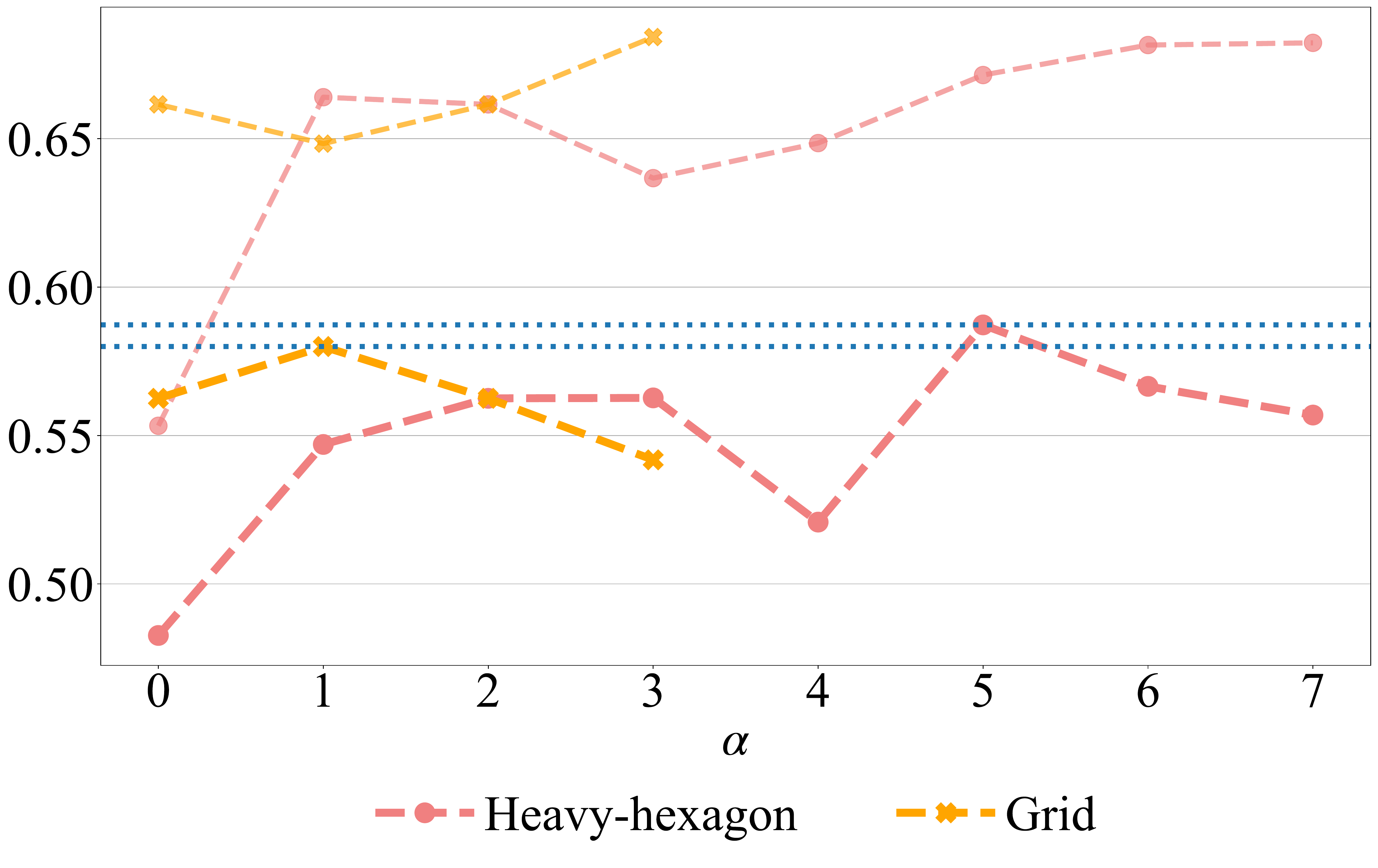}}
    \hfill
    \subfloat[QAOA-10\label{fig:qaoa_10}]
    {\includegraphics[width=0.8\linewidth]{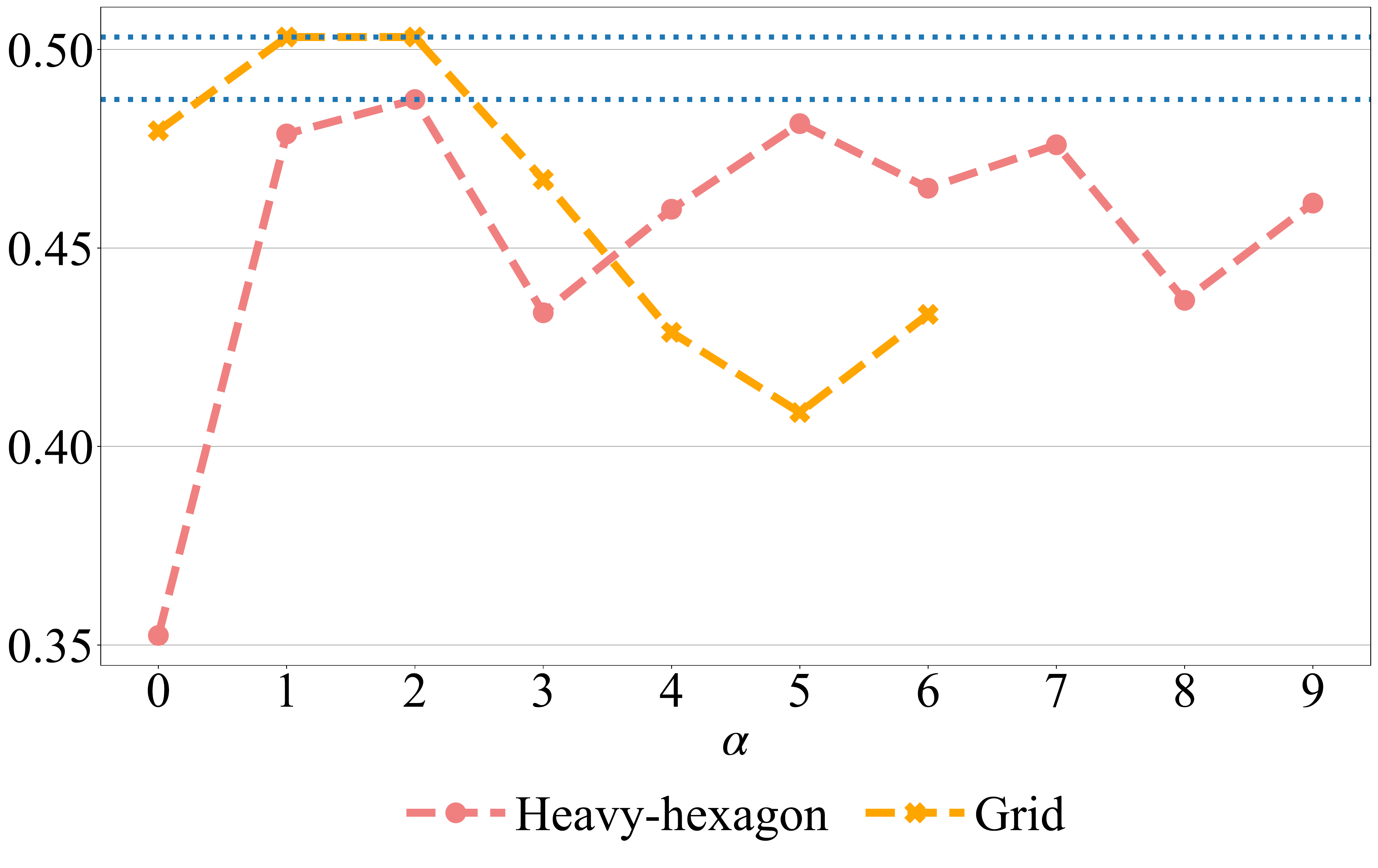}}
    \caption{Evaluation results for QAOA circuits on both architecture spaces.  Horizontal dotted lines represent the maximal fidelity values for the respective spaces.}
    \label{fig:a_qaoa}
 \end{figure}

Fig.~\ref{fig:a_qaoa} displays the evaluation results for QAOA circuits.
The results in Fig.~\ref{fig:qaoa_8} and Fig.~\ref{fig:qaoa_10} indicate that by activating two flexible edges, heavy-hexagon-based architectures can outperform the base grid architecture.
In addition, although the base grid architecture can achieve better circuit fidelity than the base heavy-hexagon architecture due to its dense qubit connectivity, architecture optimization can reduce the performance gap and can even yield a higher fidelity for heavy-hexagon-based architectures, as seen in Fig.~\ref{fig:qaoa_8}.

When ignoring crosstalk error, the fidelity for grid-based architectures is greater than or equal to that of heavy-hexagon-based architectures, which matches our expectation that higher connectivity enables more gate parallelization, reducing the number of necessary SWAP insertions.
However, with crosstalk error, high connectivity can harm the fidelity (Fig.~\ref{fig:qaoa_8}).
Therefore, we believe that crosstalk error has a significant and detrimental impact on high connectivity architectures and is the main reason that optimized heavy-hexagon-based architectures can outperform optimized grid-based counterparts.

\subsubsection{QCNN}
\begin{figure}
\centering
    {\includegraphics[width=0.8\linewidth]{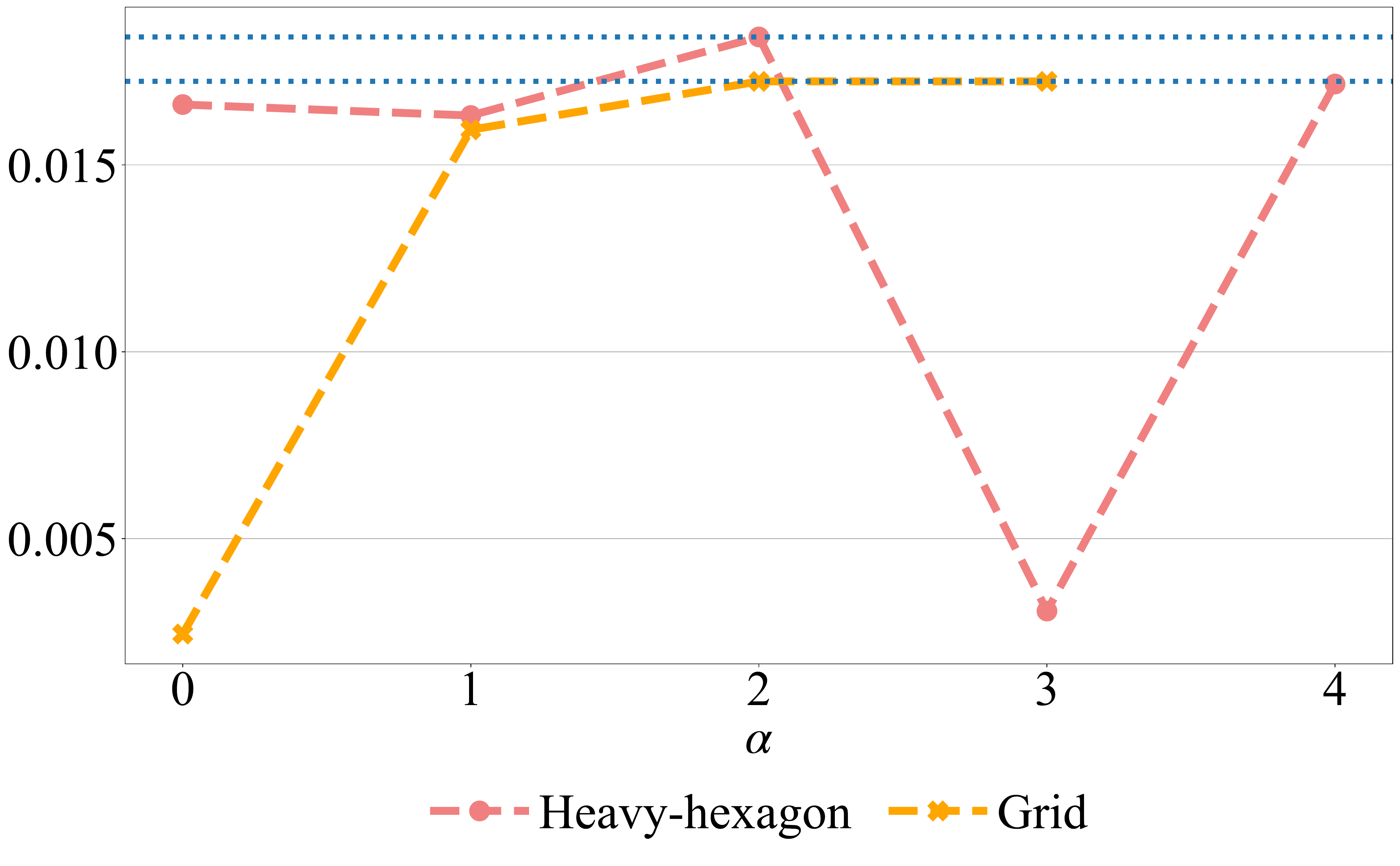}}
    \caption{Evaluation results for QCNN-8 on both architecture spaces. Horizontal dotted lines represent the maximal fidelity values for the respective space.}
    \label{fig:qcnn_8}
 \end{figure}

The evaluation results for QCNN-8 are shown in Fig.~\ref{fig:qcnn_8}.
The performance for grid-based architectures is worse than that of  heavy-hexagon-based architectures 
due to the dominating qubit idling time
and the crosstalk errors.
According to our experimental results, the average idling time is 9.118$\mathrm{\mu s}$ for the base heavy-hexagon architecture and 11.125$\mathrm{\mu s}$ for the base grid architecture, which is 1.22X longer.

\subsection{Cross Design Evaluation}
\label{subsec:cross_design_eval}
To measure the generalizability of our approach, 
we compiled different circuits of the same application onto the same optimized architecture using OLSQ to perform layout synthesis.
In this experiment, we selected the heavy-hexagon-based and grid-based architectures optimized for the QAOA-8 circuit used in Section~\ref{subsec:search_results} (target circuit) as an example for demonstration, and utilized the other thirty random QAOA graphs of size 8 for evaluation.

\begin{figure}
\centering
    
    {\includegraphics[width=0.8\linewidth]{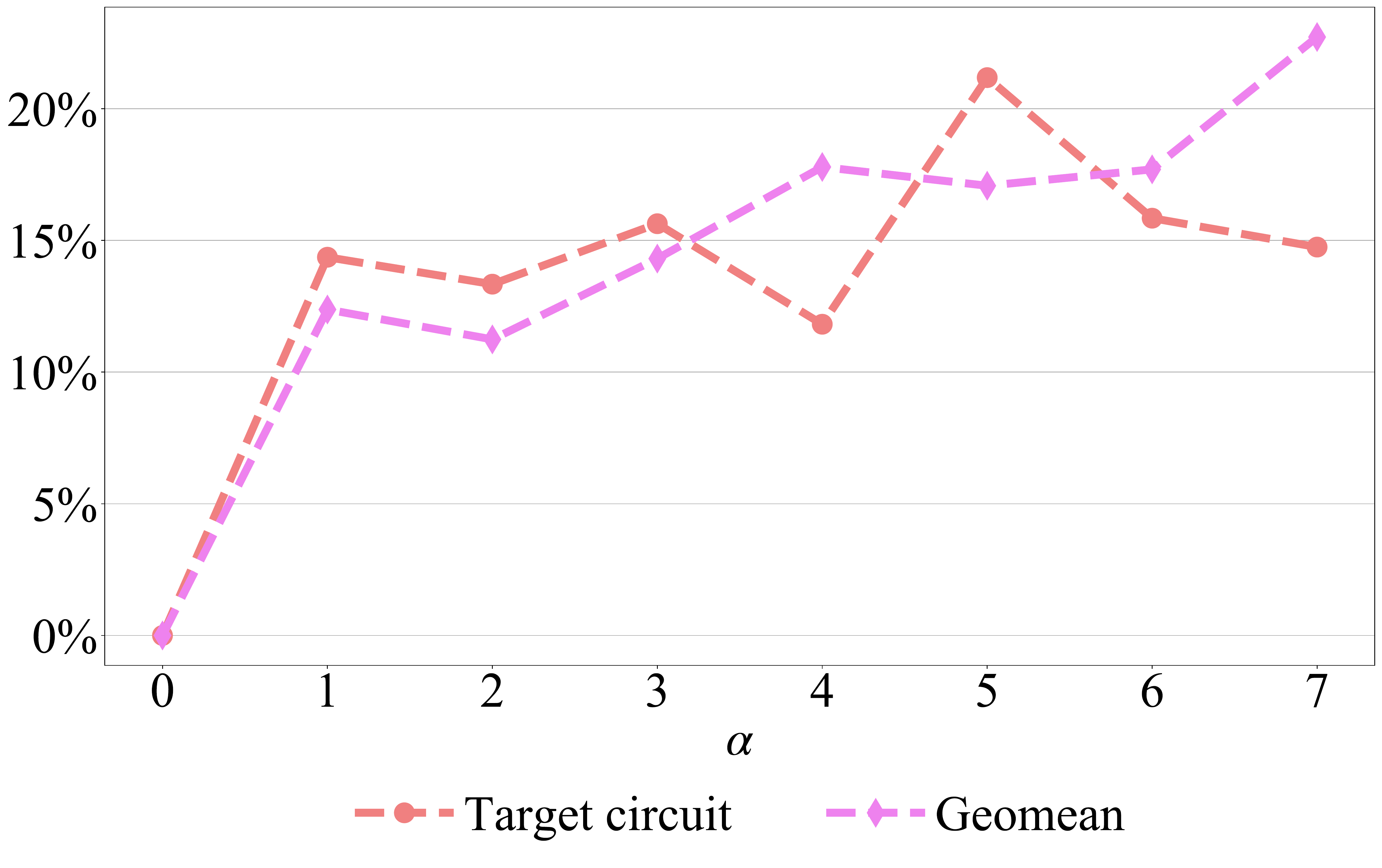}

    \caption{Cross design evaluation on the heavy-hexagon architecture space for QAOA graphs of size 8.}}
    \label{fig:cross_hh_qaoa_8}
  \end{figure}

\begin{figure}
\centering
    
    {\includegraphics[width=0.8\linewidth]{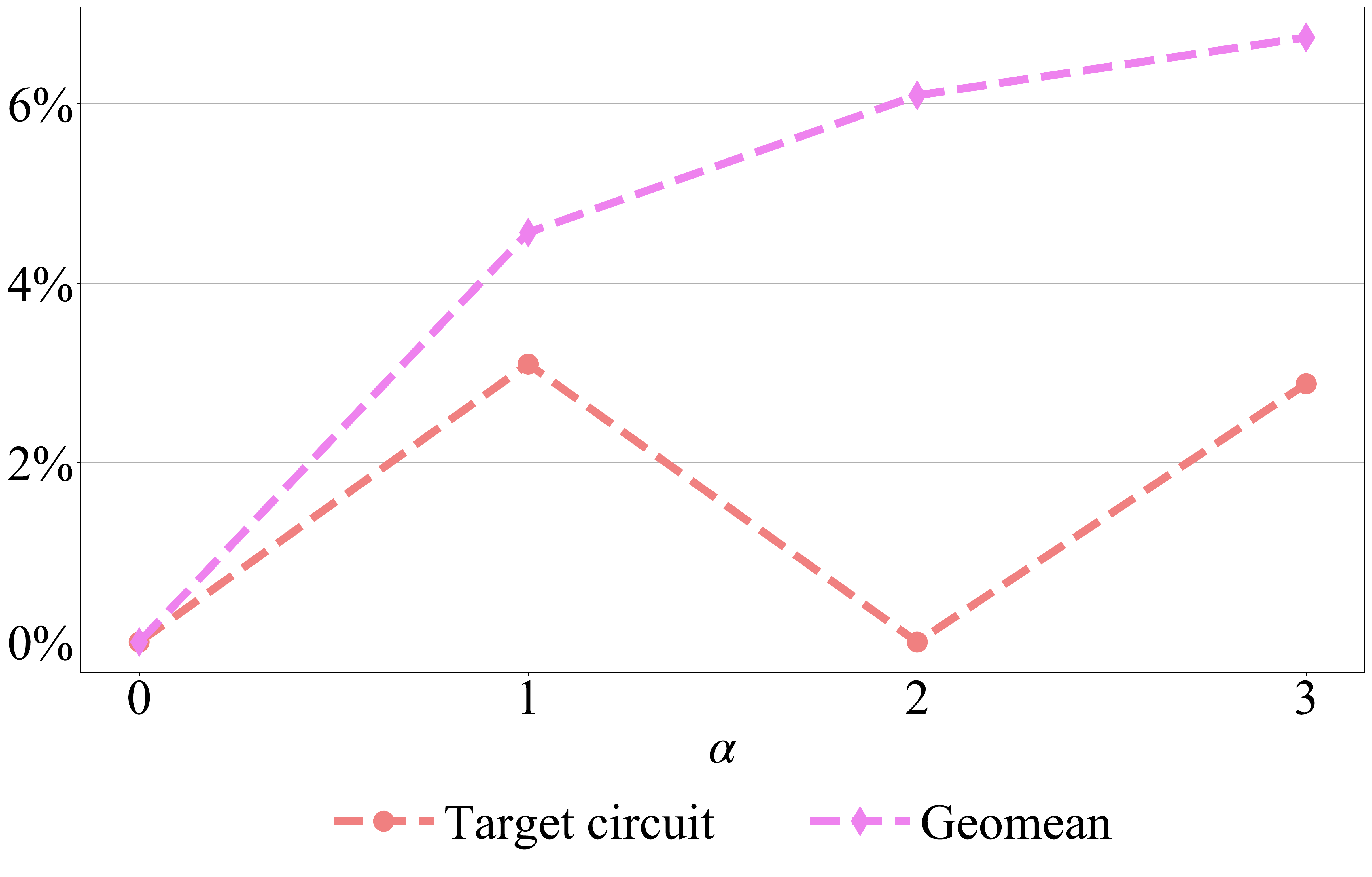}}
    \caption{Cross design evaluation on the grid architecture space for QAOA graphs of size 8.}
    \label{fig:cross_grid_qaoa_8}
  \end{figure}

The evaluation results for heavy-hexagon-based and grid-based architectures are shown in Fig.~\ref{fig:cross_hh_qaoa_8} and Fig.~\ref{fig:cross_grid_qaoa_8}, which depict the fidelity improvement for the target circuit and the average fidelity of in other 30 circuits.
The evaluation results indicate that the optimized domain-specific architectures achieve higher fidelity than that of the base architecture, i.e., the architecture with $\alpha=0$.
These results demonstrate the generality of our optimized architectures. 
Although the heavy-hexagon-based and grid-based architectures are optimized for a given circuit, the other circuits of similar type and size can also benefit from the performance gain using the optimized architectures.
In conclusion, we can achieve 22.7\% improvement on the heavy-hexagon-based architectures and 6.7\% improvement on the grid-based architectures for the non-target circuits using the optimized architectures.

%% file: 8_conclusion.tex
\section{Future Directions}
\label{sec:future}
In this study, we evaluated our methodology to search for optimal architectures for a specific quantum circuit.
Through our evaluation, we discovered that these optimized architectures can still demonstrate fidelity improvements across different instances under the same application domain.
Our future work will continue generalizing our framework to a wider range of architecture types and applications.
Rather than targeting one specific circuit for optimization, we can aim to design optimal architectures for a family of quantum circuits under a given application domain.
This would allow us to maximize the \emph{average} performance of our optimized architectures across different application instances.
In this work, we have benchmarked our framework on quantum optimization algorithm and quantum machine learning.
Our framework are not restricted to these domains and can be also applied to improve the performance of variational quantum algorithms~\cite{peruzzo2014vqe} and quantum chemistry~\cite{aspuru2005simulated_quan_chemisty}.
Additionally, we can test our optimized architectures on the application-oriented benchmarks~\cite{lubinski2021application_benchmark,tomesh2022supermarq}, which provides a great performance evaluation for domain-specific architectures.

We can generalize our current approach to different application domains through multi-objective optimization while providing performance guarantees to each target domain.
This is relevant for practical application, in which only a limited number of architectures can be realized due to fabrication overhead.
When generalizing our framework to take a set of circuits as inputs, we can assign independent compilers from a portfolio of compilers to each circuit instance while sharing architecture parameters for joint optimization.
Combining our current framework with the aforementioned optimization schemes will allow us to develop a two-stage  architecture optimization flow.
In the first stage, the domain-specific architecture optimization, our goal would be to design the lower-level architecture, such as the locations of tunable coupling edges. 
The optimization in this stage focuses on improving the architecture performance towards a family of application domains.
In the second stage, the instance-specific architecture optimization, our goal would be designing higher-level, or control-level architecture, such as which coupling edges to activate for which parts of a quantum circuit.
Here, the optimization targets a specific quantum circuit of interest and determines the architecture parameters that can change depending on the quantum controls.
With this two-stage design, we can fully utilize the power of quantum computing by achieving high circuit fidelity on a fully customizable architecture.

An orthogonal direction is to improve the scalability of our SMT-based approach.
Our current implementation can be applied to optmize near-term quantum architectures by directly taking the whole layout as input or adopting the tiling strategy, which focuses on optimizing a tiled layout with the goal to design the optimal patterns that can be repeated to constitute the larger layout. 
With either approach, it would be beneficial to further improve the scalability.
One possible acceleration method is to utilize different encoding methods, e.g., uninterpreted functions and bit-vectors, which have been shown can achieve significant speedup~\cite{nikolaj2021plant_z3}.

\section{Conclusion}
\label{sec:conclusion}

In this paper, we proposed and implemented the first domain-specific quantum architecture optimization framework that integrates quantum circuit compilation into architecture optimization in order to improve the fidelity of quantum applications. 
We formulated the architecture optimization as a constraint satisfication problem to guarantee the output performance.
Under this framework, we presented an algorithm to search for the optimal architecture for circuit compilation under hardware-specific constraints.
To efficiently evaluate the given quantum architecture, we proposed a quantum architecture evaluation flow that approximates exact circuit fidelity with a linear fidelity cost function.
More specifically, we include an analytical crosstalk error model to account for realistic performance impairment due to quantum control crosstalks.
Our evaluation shows 59\% fidelity improvement on average for a 10-qubit QAOA MAXCUT circuit on the heavy-hexagon-based architecture and a 14\% improvement on the grid-based architecture. 
For an 8-qubit QCNN circuit, architecture optimization improves fidelity by 11\% on the heavy-hexagon-based architecture and 605\% on the grid-based architecture.
These results demonstrate the large amount of untapped potential in traditional quantum architecture, which can be taken advantage of under our domain-specific architecture optimization framework.
For example, the heavy hexagon architecture is commonly considered to be unfriendly for NISQ application due to its lack of connectivity, but with even the modest modification of adding two flexible coupling edges, a significant fidelity improvement can be achieved for the QAOA application.

Our results showed that under a  scalable and automated  design and evaluation flow, the crafted quantum architectures can be superior to the manually-designed architectures for a given application domain of interest.
In addition, our proposed architecture optimization framework is general and versatile.
Although we focused on the superconducting qubit, this framework is directly applicable to other types of qubits, e.g., the trapped-ion qubit and neutral-atom qubits.
Moreover, the optimization objectives in our architecture optimization flow can be adapted to account for changing requirements in quantum applications. Lastly, the definition of architecture space can be extended beyond a qubit coupling graph. 
For example, our architecture space can contain not only the adjustable couplings between a fixed number of qubits, but the qubit number as well.
The domain-specific architecture optimization framework proposed in this work opens technology pathways for an automated architecture design process and provides the essential toolbox for research efforts in developing large-scale quantum hardware for both general and special purpose quantum computation.

%% file: biography.tex
\begin{IEEEbiography}[{\includegraphics[width=1in,height=1.25in,clip,keepaspectratio]{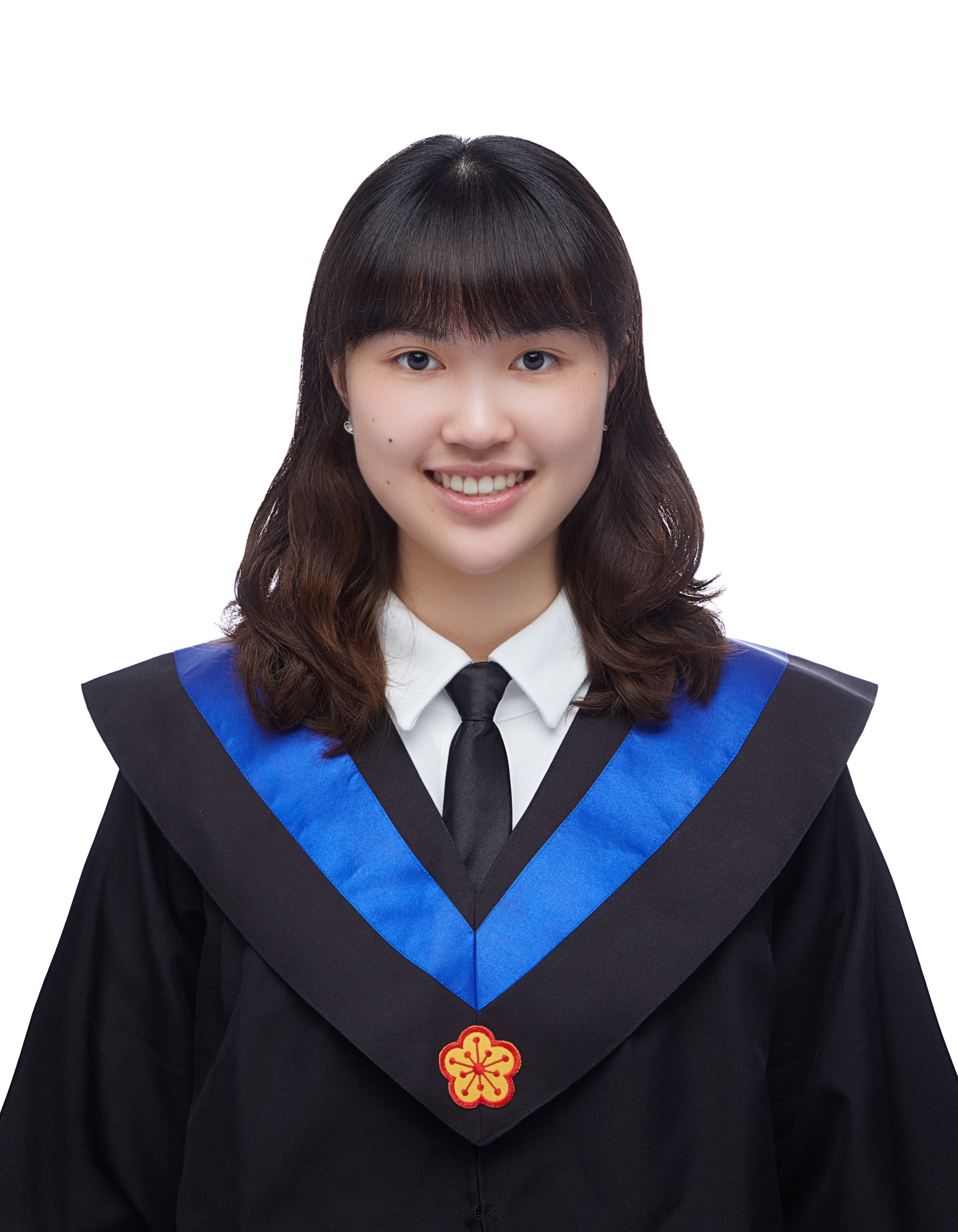}}]
{Wan-Hsuan Lin}
received a B.S. in Electrical Engineering from National Taiwan University (NTU), Taipei, Taiwan, in 2021. She is currently pursuing a Ph.D. degree in Computer Science with the University of California, Los Angeles, CA, USA. Her current research interest focuses on design automation for quantum computing.
\end{IEEEbiography}

\begin{IEEEbiography}[{\includegraphics[width=0.9in,height=1.2in,clip,keepaspectratio]{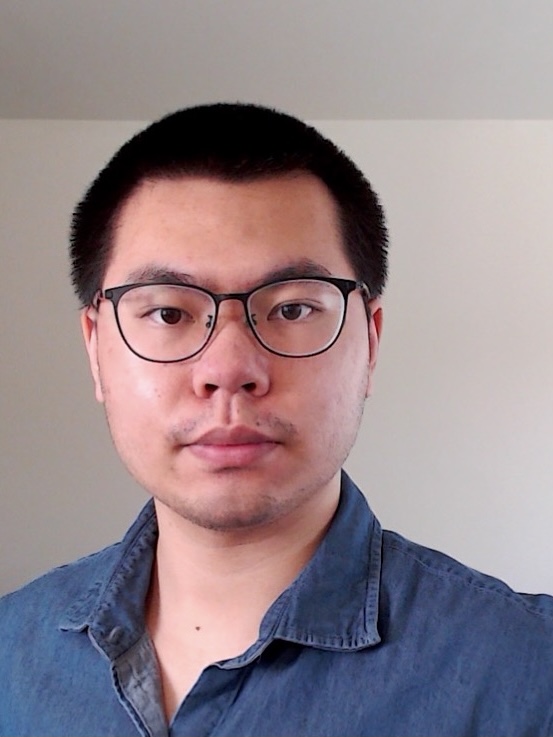}}]
{Bochen Tan}
received the B.S. degree in electrical engineering from Peking University in 2019, and the M.S. degree in computer science from University of California, Los Angeles in 2022.
He is currently a graduate student researcher at UCLA focusing on design automation for quantum computing.
\end{IEEEbiography}

\begin{IEEEbiography}
[{\includegraphics[width=1in,height=1.25in,clip,keepaspectratio]{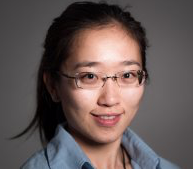}}]
{Murphy Yuezhen Niu}
 is a senior research scientist in Google Quantum AI team. She obtained Ph.D. in physics from MIT, and B.S. in physics from Peking University. Her research has focused on applying machine learning methods to quantum optimization, calibration, system learning, and quantum algorithm designs. Murphy leads the theoretical effort in modeling, calibration, mitigation, and simulation of correlated errors in large-scale superconducting qubit systems at Google. Murphy is the recipient of Claude E. Shannon Research Award from MIT RLE.
\end{IEEEbiography}

\begin{IEEEbiography}
[{\includegraphics[width=1in,height=1in,clip,keepaspectratio]{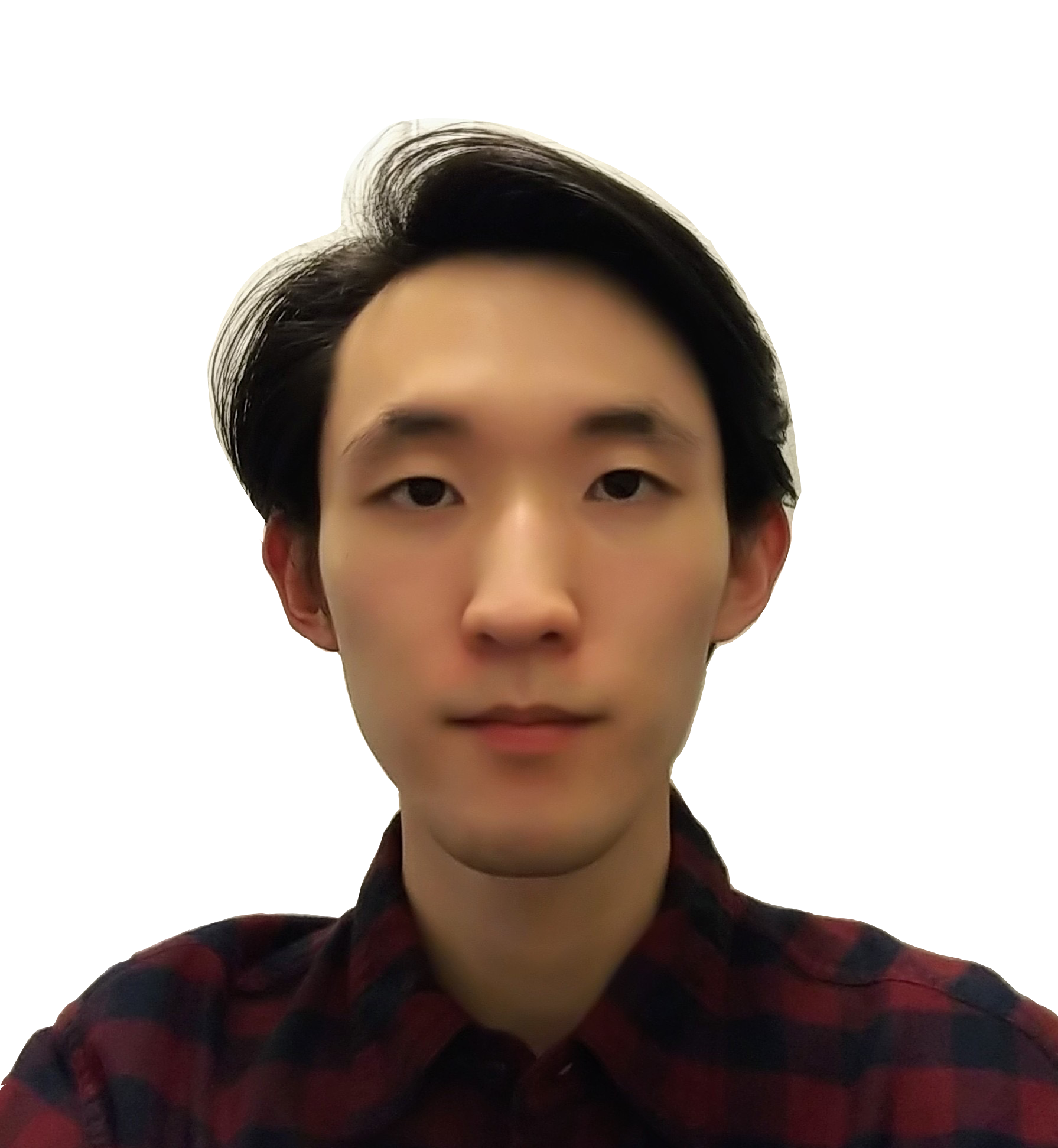}}]
{Jason Kimko}
received his B.S. in Computer Science and Mathematics from the College of William \& Mary in 2018.
He has previously helped maintain and develop a massively parallel, multiphysics simulation code at Lawrence Livermore National Laboratory.
Currently, he is pursuing a Ph.D. in Computer Science at the University of California, Los Angeles.
His research interests lie in the development of performant codes through parallelization and hardware acceleration.
\end{IEEEbiography}

\begin{IEEEbiography}
[{\includegraphics[width=1in,height=1.25in,clip,keepaspectratio]{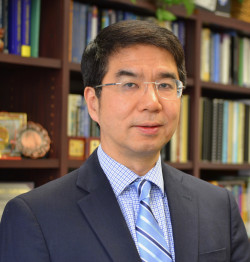}}]
{Jason Cong (Fellow, IEEE)}
received a BS degree in Computer Science from Peking University, Beijing, China, in 1985, and bath an MS and a PhD in Computer Science from the University of Illinois at Urbana-Champaign, Champaign, Illinois, in 1987 and 1990, respectively. Currently, he is a   (and a former chair) with the UCLA Computer Science Department, with a joint appointment from the Electrical Engineering Department, the director of Center for Domain-Specific Computing (CDSC),
and the director of VLSI Architecture, Synthesis, and Technology (VAST) Laboratory. His research interests include novel architectures and compilation for customizable computing, design automation for VLSI systems and other emerging technologies, such as quantum computing and highly scalable algorithms. He has more than 500 publications in these areas, including 16 best paper awards, three ten-year most influential paper awards, and one paper inducted into the FPGA and Reconfigurable Computing Hall of Fame. He was elected to an ACM fellow in 2008 and a member of the National Academy of Engineering in 2017.
\end{IEEEbiography}